\documentclass[11pt]{article}

\usepackage[a4paper,lmargin=2.5cm,rmargin=2cm,tmargin=1cm,bmargin=1cm,includehead,includefoot]{geometry}

\usepackage{graphicx,amsmath,url}
\usepackage{microtype,lmodern}
\usepackage{mathtools,amssymb,gensymb}
\usepackage{tgtermes}
\usepackage[T1]{fontenc}
\usepackage[utf8]{inputenc}

\usepackage{acronym}
\usepackage{amsmath}
\usepackage{array}
\usepackage{bbold}
\usepackage{booktabs}
\usepackage{dsfont}
\usepackage{fancyhdr}
\usepackage{relsize}
\usepackage[font={small,up,singlespacing}]{subcaption}
\usepackage[backend=bibtex,style=ieee,sorting=none]{biblatex}
\usepackage{pdfpages}
\usepackage{bbm}
\usepackage{color}
\usepackage{siunitx}
\usepackage{braket}
\usepackage{mwe}
\usepackage{biblatex}
\usepackage{multirow}
\usepackage{mathtools}

\usepackage[colorlinks,breaklinks]{hyperref} 
\usepackage{cleveref}

\DeclarePairedDelimiter\ceil{\lceil}{\rceil}

\hypersetup{hypertexnames=true, linkcolor=blue, anchorcolor=black, citecolor=blue, urlcolor=blue}
\graphicspath{{./}{imglocal/}{img/}}
\title{\textbf{Regularized by Physics: Graph Neural Network Parametrized Potentials for the Description of Intermolecular Interactions}}

\author{Moritz Th\"urlemann, Lennard B\"oselt, Sereina Riniker*}
\date{Laboratory of Physical Chemistry, ETH Zurich, Vladimir-Prelog-Weg 2, 8093 Zurich, Switzerland. Email: sriniker@ethz.ch}

\begin{document}
\pagestyle{fancy}
\rhead{}
\lhead{MLFF}
\chead{}
\maketitle

\section{Abstract}
Simulations with an explicit description of intermolecular forces using electronic structure methods are still not feasible for many systems of interest. As a result, empirical methods such as force fields (FF) have become an established tool for the simulation of large and complex molecular systems.
However, the parametrization of FF is time consuming and has traditionally been based largely on experimental data, which is scarce for many functional groups. Recent years have therefore seen increasing efforts to automatize FF parametrization and a move towards FF fitted against quantum-mechanical reference data.
Here, we propose an alternative strategy to parametrize intermolecular interactions, which makes use of machine learning and gradient-descent based optimization while retaining a functional form founded in physics. This strategy can be viewed as generalization of existing FF parametrization methods.
In the proposed approach, graph neural networks are used in conjunction with automatic differentiation to parametrize physically motivated models to potential-energy surfaces, enabling full automatization and broad applicability in chemical space. 
As a result, highly accurate FF models are obtained which retain the computational efficiency, interpretability and robustness of classical FF. 
To showcase the potential of the proposed method, both a fixed-charge model and a polarizable model are parametrized for intermolecular interactions and applied to a wide range of systems including dimer dissociation curves and condensed-phase systems.

\section{Introduction}
Computer based simulations are a powerful tool for the investigation of chemical systems \cite{DEShawFolding, BiomolecularModeling, Warshel1976}. 
Performing such simulations requires an accurate description of intermolecular forces \cite{Stone}. 
However, due to the computational complexity of \textit{ab initio} methods \cite{Schrodinger, Hartree, Fock, Slater, MP2} or density functional theory (DFT) \cite{DFT1, DFT2}, an exact description is out of reach for most systems, particularly systems of biological relevance \cite{QuantumComplexity, QMScaling}.
As a result, more approximate methods have been developed, which can be broadly categorized into three classes: 
semi-empirical methods, classical force fields (FF), and machine learning (ML) based models.

Semi-empirical methods explicitly describe the electronic structure \cite{Levine2013, PopleSemiEmpirical}. However, various approximations are introduced to reduce computational costs.
Existing methods attempt to compensate for these approximations by introducing a relatively small number of empirical parameters \cite{PM7, DFTB, XTB}. 
Classical force fields, on the other hand, forego an explicit description of the electronic structure and employ instead a predefined functional form and associated parameters, which together encapsulate aspects of a given interaction \cite{FixedChargeFF, PolarizableFF}. 
Due to their simplicity, they can be evaluated much more efficiently than \textit{ab initio} or semi-empirical methods, but generally require a larger number of parameters and extensive parametrization.
In recent years, ML-based models have emerged as a third alternative \cite{MLPotentials}, which assume fewer inductive biases, but require an even larger number of parameters compared to the two previous approaches.
Even though very promising results have been reported for ML potentials \cite{MLPotentials, GDML, SCHNET, DIME, SOAP, PAINN}, their application to condensed-phase systems and the prediction of experimentally measured properties has been fairly limited \cite{MrBoeselt, ANI2, PhysNet, GDMLCoarse, MLCondensedPhaseBehler, SIMPLENN}.

The relative scarcity of application of ML potentials to propagate molecular dynamics (MD) simulations is likely a result of insufficient data efficiency due to a lack of inductive biases as well as difficulties posed by the presence of a large number of relatively weak and long-ranged interactions in condensed-phase systems \cite{FailuresShortRangeML}. 
However, there are also discussions whether ML potentials describe features of the potential-energy surface (PES), such as its curvature, sufficiently accurate to perform MD simulations \cite{MLBonds}.
With their predefined functional form, which is physically motivated, FF have become an established tool to simulate large condensed-phase systems such as solvated proteins over long time-scales \cite{DEShawFolding}. 
While the FF formalism provides a computationally efficient, interpretable and robust way to describe forces in molecular systems, this robustness comes at a price. 
Most of the commonly used FF do not account for phenomena such as charge anisotropy or polarization.
In addition, the development of a FF is still a non-trivial process, despite advances in automation over the past years \cite{Paramol, ForceBalance1, ForceBalance2, TAFFI, OnTheFlyParametrization, CombiFF, AmoebaML}. The OpenFF initiative in particular has initiated a grand effort to fully automate this process, including atom-typing \cite{AutomaticTyping, ChemicalPerception}, data generation \cite{TorsionScanning}, parametrization \cite{ParsleyFF}, and validation \cite{AutomatedFFBenchmarking}.

While FF were historically and continue to be (partly) fitted to experimental data, the advances in computational power and improved scaling of methods based on quantum-mechanical (QM) calculations has opened up new opportunities \cite{FixedChargeFF}.
As a readily accessible data source, focus has shifted to parametrization with respect to QM reference data such as torsion profiles or interaction components.
In addition, there is an increased effort to extract FF parameters directly from electron densities \cite{QUBEKit, AIMFF, MEDFF, QMDispersionFF, FFFromSAPT, FFFirstPrinciples, QMDFF}. 
As an example, van Vleet \textit{et al.} \cite{BeyondBornMayer} developed an \textit{ab initio} FF based on a parametrization formalism using Slater functions from which certain parameters were directly derived.
An alternative approach is to keep a predefined functional form to describe intermolecular potentials but obtain the parameters from a ML model \cite{PhysicsBasedML, CLIFF}. 
In a similar fashion, Wang \textit{et al.} \cite{ChoderaEndToEnd} and Harris \textit{et al.} \cite{HarrisEndToEnd} investigated graph neural networks (GNN) and graph-convolutional neural networks (GCNN) in combination with automatic differentiation as a method to parametrize FF.
Focusing on intramolecular interactions, they could show that GNN can be used to predict FF parameters from potential energies and recover human-defined atom types.
Finally, we also point out recent efforts which use ML in a complementary fashion to extract symbolic expressions from data \cite{RennerSymbolic, SymbolicPotential, SymbolicDynamical, SymbolicTegmark}.

In this work, we build on these developments to propose a universal framework for the parametrization of FF, focusing on intermolecular interactions.
Besides generalization of the parametrization process, we describe a formalism for end-to-end differentiable FF, taking advantage of learned atom types. 
The proposed approach is applied to the parametrization of a non-polarizable as well as a polarizable FF. 
Multipoles and monopoles used to describe electrostatic interactions were obtained separately from our previously introduced equivariant GNN model without further modifications \cite{MultipolesMe}. 
For both the non-polarizbale and polarizable FF, the model is trained on the PES of dimer interaction potentials of the recently published DES5M data set \cite{DEShawDimers} and a data set of intermolecular potentials of molecular crystals, which was generated for this work.
Hence, parameters are learned by the model from scratch to reproduce the given PES.
We show that a fixed-charge FF parametrized in such a manner can be used to successfully reproduce experimental condensed-phase properties.
We find that the resulting models outperform comparable models for a wide range of test cases.

The work is structured as follows: In the Theory section, the proposed formalism and potential energy terms are introduced. In the Methods section, the training and validation procedures is outlined. Finally in the Results and Discussion section, the performance of the models on a wide range of test systems and properties are discussed.

\section{Theory}
\subsection{Formalism}\label{sec:formalism}
Assuming a chemist's viewpoint, molecules can be interpreted as graphs $G=(A, B)$ with nodes (atoms) $A$ and edges (bonds) $B$.
Accordingly, a FF consists of a function $\mathcal{G}: G \rightarrow \theta$,
which assigns FF parameters $\theta$ to a molecular graph $G$ and a functional form
\begin{equation}
    \mathcal{V}_\theta (x) = \sum_i \mathcal{V}_{\theta, i}(x)
\end{equation}
where the total potential energy of a state $x$ is decomposed into $\mathcal{V}_{\theta, i}$. 
The function $\mathcal{G}$ that assigns parameters $\theta$ can be understood as a parametrized, or learnable, function itself. 
For commonly used FF, $\mathcal{G}$ is generally expert-devised and depends only on atomic features such as element types and hybridization states. For FF that are parametrized based on electron densities, $\mathcal{G}$ partitions and assigns parameters to an electron density.

\begin{figure}
    \centering
    \includegraphics[width=0.99\textwidth]{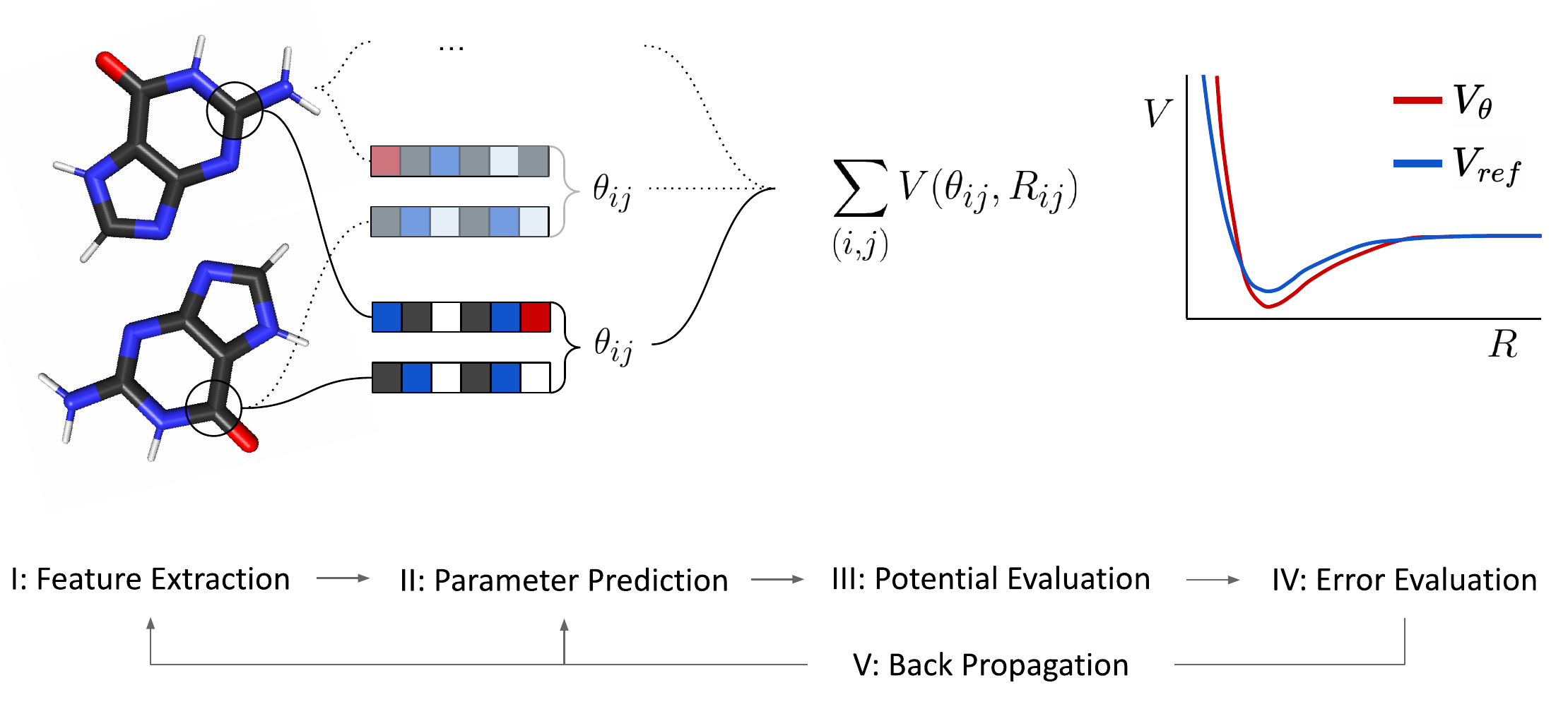}
    \caption{Overview of the proposed parametrization workflow: I) Atomic environments are encoded as feature vectors with a GNN. II) Parameters $\theta_{ij}$ are predicted for tuples of atomic features. III) In conjunction with predefined parametric interaction potentials $V$, the predicted parameters $\theta_{ij}$ are used to evaluate the potential energy of a state. IV) The prediction is evaluated against reference properties. V) Using automatic differentiation, errors with respect to the target properties are backpropagated to improve the quality of the predicted parameters.}
    \label{fig:overview}
\end{figure}

Given a parametrization function $\mathcal{G}$, a FF can thus be interpreted as a function $\mathcal{V_\theta}$ with parameters $\theta$ and a functional form $\mathcal{V}$, $\mathcal{V}_\theta : X \rightarrow V$,
which maps a PES $V\in\mathbf{R}$ to the states $X\in\mathbf{R}^n$ of a system, with $n$ denoting the dimensionality of the system.
A system can thus be propagated in time by integrating the negative derivative of the potential energy $v\in V$ with respect to its current state $x \in X$, 
\begin{equation}
    \mathcal{F}(x) = -\nabla_xv = -\nabla_x\mathcal{V}_\theta(x) ,
\end{equation}
where $F\in\mathbf{R}^n$ is the gradient field of the negative potential energy.
Propagating the system for an appropriate amount of time, a system property $P$ can be derived as $\mathcal{P}: X \rightarrow P$
with a function $\mathcal{P}$, which assigns a property to a state or set of states of the system.
Generally, system properties can be scalar, vectorial, or tensorial. They may be defined for each configuration or ensemble average, and depend on the functional form of the FF and its parametrization $\mathcal{V}_\theta$ (Figure \ref{fig:overview}).
Given a system property obtained from a simulation, $P_{\text{pred}}$ and a reference value $P_{\text{ref}}$ (e.g., from an experiment or QM reference calculation), a loss $L$ can be defined as $\mathcal{L}: (P_{\text{pred}}, P_{\text{ref}}) \rightarrow L$ with a loss function $\mathcal{L}$. 
Due to its generality, any computable property can be used as a target. Examples include the potential energy, gradients or Hessians from QM reference calculations, but also experimental properties would be possible such as geometrical constraints from NMR or crystallography, vibrational spectra or ensemble properties such as enthalpies of phase transitions.

\subsection{Parametrization}
Given a loss function $\mathcal{L}$ and parameters $\theta$, a FF may be optimized to yield more accurate properties. Specifically, the derivative of the loss function with respect to the FF parameters,
\begin{equation}
    \frac{\partial \mathcal{L}}{\partial \theta} = \frac{\partial \mathcal{L}}{\partial \mathcal{P}_\theta}\dots\frac{\partial \phi}{\partial \theta} ,
\end{equation}
can be used to update FF parameters successively with gradient descent based optimization methods analogously to backpropagation used in deep learning \cite{DeepLearning}. 
With the help of automatic differentiation libraries, these gradients can be evaluated with minimal additional programming effort and computational cost \cite{TensorflowPaper, PytorchAD}.

\subsection{Graph Neural Networks as Universal Parametrization Functions}
GNN are ML models parametrized by artificial neural networks (ANN) that process graph-structured data. In the commonly used form, node, edge and/or global features are iteratively refined based on the current features. GNN models differ mainly by the features used, the way the underlying graph is constructed as well as the updating or feature-refinement process applied  \cite{GNN0, GNN, GilmerQuantumGNN, InteractionNetwork}.

Considering a molecular graph $G=(A, B)$ with nodes $a_i \in A$ and edges $b_{ij} \in B$, message passing can be defined as
\begin{equation}
    \begin{aligned}
        m_{ij} &= \phi_b (h_i^l, h_j^l, b_{ij})\\
        m_i &= \sum_{j\in N(i)}m_{ij}\\
        h_i^{l+1} &= \phi_h(h_i^l, m_i) ,
    \end{aligned}
\end{equation}
with $h_i^l\in \mathbf{R}^n$ describing the hidden feature vector of node $a_i$ after $l$ graph processing steps, $b_{ij}\in \mathbf{R}^n$ the bond feature of bond $b_{ij}$ between node $i$ and $j$, $N(i)$ denoting the set of neighbours of $a_i$ and $m_{ij}$ the message from node $j$ to node $i$. $\phi_b$ and $\phi_h$ represent ANN parametrized edge and node update functions. The superscript $l$ denotes the current layer or the current iteration in the recurrent realization. 
After $n$ iterations, the refined node feature $h_i^n$ is used as an atomic-environment descriptor in subsequent steps. We note that $h_i^n$ does not necessarily have to be obtained from a GNN. In principle, any other descriptor can be used, for instance atom-based topological fingerprints.
However, GNN present themselves as a natural choice to work with graph-structured data, which in turn is ideally suited for a classical description of molecules.

\subsection{Parameter Prediction}
To predict atomic parameters, learned atom features $h_i^n$ obtained from the GNN serve as descriptors of the atomic environment. Assuming that FF parameters are a function of the atomic environment, such features can be used to distinguish atom types and to assign FF parameters.
Hence, atomic parameters are predicted as
\begin{equation}
    \theta_{i} = \phi_{\text{atom}}(h_i^n) ,
\end{equation}
with $\phi_{\text{atom}}$ denoting the readout function that assigns the parameter $\theta_i$ for a given interaction to an atom type.
In general, FF also describe how parameters of two particles are combined to describe a given interaction. Standard biomolecular FF use combination rules to derive parameters of two distinct atom types \cite{FixedChargeFF}. By design, combination rules must be symmetric under arbitrary permutations of types present in the interaction.
Instead of using established combination rules such as arithmetic or geometric means, pairwise parameters are predicted as a function of two atom types.
Specifically, the following approach was chosen
\begin{equation}
    \label{eq:combination_rule_1}
    \theta_{ij} = \phi_{\text{pair}}(h_i^n, h_j^n) + \phi_{\text{pair}}(h_j^n, h_i^n)
\end{equation}
In this case, $\phi_{\text{pair}}$ is an ANN parametrized function, which assigns the parameters of a specific interaction to a pair of atomic environments.
This approach guarantees symmetry with respect to permutation of two atoms and allows for more complex combination rules.

\subsection{Force-Field Parametrization Function}
Combining the previously introduced concepts with the formalism described in Section \ref{sec:formalism}, the FF parametrization $\mathcal{G_\phi}$ can be defined as the combination of the following two components: (i) a typing function to assign atom types to a given system, and (ii) a combination rule, which returns parameters for a set of atoms partaking in a given interaction.
In the current work, the function assigning atom types is modelled with a GNN consisting of a node and edge update layer, $\phi_h$ and $\phi_b$, and a combination rule $\phi_{\text{Atom}}$ or $\phi_{\text{Pair}}$ for atomic and pairwise parameters, respectively.
In the context of the GNN formalism, $\phi_\theta$ can be considered a readout function, which maps the hidden state of node features to a label. In our case, FF parameters are mapped to interactions between given atom types.
Evidently, established FF atom type definitions and parametrization procedures can be viewed as a special case of the described formalism.
For example, the SMARTS patterns used for chemical perception in Open FF \cite{ChemicalPerception} can be cast as graph-based operations that account for features such as the element, its coordination number, bonded neighbours, or sub-graph features. In the case of models, which derive parameters from electron densities, the partition function used to decompose the electron density into atomic contributions takes up the role of the typing functions. In these cases, the combination rules are often derived from first principle considerations or empirically fitted \cite{QUBEKit, MEDFF}.

\subsection{Models}
Besides the aforementioned components, which assign parameters to a given interaction or atom, a FF must further define potential-energy terms and an associated functional form. 
The chosen functional form reflects the assumptions of the model, and thus determines the accuracy level of the model, its computational cost, and its capabilities.

In this work, we consider two models. The first model is based on three interactions: (i) repulsive, (ii) attractive, and (iii) electrostatic. The first two components follow the functional form of the Mie potential \cite{Mie} with a repulsive $C_9$ term and an attractive $C_6$ term. The electrostatic component is described with Coulomb's law and fixed partial charges.
Hence, this model assumes an isotropic-pairwise-additive form, and we will refer to it as ``IPA model''.
The second model considered in this work is based on multipole electrostatics and polarizable atoms, therefore violating isotropy and pairwise additivity. We will refer to this anisotropic-non-additive model as ``ANA model''.

The IPA model is similar to the functional form found in the most widely used FF \cite{FixedChargeFF, GAFF, 99SB, OPLS, GROMOS1, CHARMM, ParsleyFF}. 
The functional form of the ANA model, on the other hand, is an attempt to explore the limit of a purely classical model through an implicit description of the electron distribution based on atomic multipoles and induced dipoles. Its functional form is similar to the class of polarizable and QM derived FF \cite{AMOEBA, AMOEBA+, AMOEBAChargeDamping, HIPPO, AnisotropicExchange, BJ, CLIFF, PhysicsBasedML}.
In the following section, the components used in the respective models are described.

For consistency, the following notation is used: Capitalized letters refer to pairwise parameters, i.e., parameters given for a pair of atoms. Small letters are used for atomic parameters, i.e., parameters assigned to one specific atom. Subscripts are used to further clarify interaction partners.
In general, the indices $i, j$ iterate over each unique pair of atoms.

\subsection{Potential-Energy Terms: IPA Model}
The IPA model for the non-bonded potential energy includes three components: attractive-repulsive described with the Mie potential \cite{Mie}, and the electrostatic interaction described with atomic partial charges interacting through Coulomb's law,
\begin{equation}
    V^\text{pot,IPA} = V^\text{Mie} + V^\text{ele}
\end{equation}

\subsubsection{Attraction-Repulsion Potential}
As the simplest potential with two parameters that can reproduce the qualitative features of the dissociation of uncharged atoms, the Mie potential \cite{Mie} is used,
\begin{equation}
    V^\text{Mie}_{ij}(r_{ij})=\frac{C_n(i,j)}{r^n_{ij}} - \frac{C_m(i,j)}{r^m_{ij}} ,
\end{equation}
where $C_n$ and $C_m$ denote the coefficients used to describe the strength of the repulsion and the attraction, respectively, and $r_{ij}$ the distance between two atoms $i$ and $j$. 
For $n=12$ and $m=6$, the well-known Lennard-Jones potential is obtained \cite{LJ1}.
While the attractive part is often set to $m=6$ motivated by the asymptotic behaviour of the dispersion interaction, there is no such obvious choice for the repulsive part. 
In the past, $n=12$ was commonly chosen for its computational efficiency.
In this work, a softer $n=9$ repulsive interaction is used as in Ref.~\cite{COMPASS}. This choice allows for a more accurate description of the repulsive interaction while still retaining computational efficiency and comparability with the more common $C_{12}-C_{6}$ formulation. 

\subsubsection{Electrostatics}
Electrostatic interactions in the IPA model are treated on the basis of atomic monopoles (partial charges),
\begin{equation}
    V^\text{ele}_{ij}(r_{ij})=\frac{1}{4\pi\epsilon_0}\frac{q_iq_j}{r_{ij}} 
\end{equation}
with monopoles $q_i$ and $q_j$, and the vacuum permittivity $\epsilon_0$.
In this work, the monopoles are obtained from our previously introduced equivariant GNN model \cite{MultipolesMe}, which was trained on minimal basis iterative Stockholder multipoles (MBIS) \cite{MBIS}.

\subsubsection{Predicted Parameters}
The IPA model predicts a set of $C_6$ and $C_9$ parameters for each atom pair, whose features were represented with the permutation-invariant pairwise feature combination shown in Eq.~(\ref{eq:combination_rule_1}). No prior knowledge such as baseline or default parameters is used in the training of the IPA model.
As mentioned above, the monopoles $q_i$ are obtained from a separate GNN model \cite{MultipolesMe} and not further modified for the present work.

The atomic monopoles and the parameters of the Mie potential are considered fixed parameters, i.e., they remain constant during a simulation and do not change in response to changes in the molecular geometry. This setting was chosen to be comparable with existing FF \cite{FixedChargeFF}.

\subsection{Potential-Energy Terms: ANA Model}
Unlike the IPA model, the second model includes explicit treatment of polarization effects and anisotropy, resulting in an anisotropic and non-additive model, abbreviated as 'ANA' model.
The model is motivated by the desire to develop a fully classical description that performs with an accuracy expected from semi-empirical methods.
In addition, we aim to demonstrate the power of the proposed parametrization strategy through its application to a model with several inter-dependent components.
At its core, the ANA model is based on an implicit description of the electronic structure through the use of atomic multipoles and a polarization model.
Adding additional interaction terms allows for a more detailed decomposition of the total energy. Consistent with the decomposition by symmetry adapted perturbation theory (SAPT) \cite{SAPT1}, components for the dispersion, electrostatic, induction, and exchange potential energy are used. As a further benefit, it is possible to include SAPT terms in the fitting procedure.

The functional form of the non-bonded potential energy of the ANA model is inspired by previous work on polarizable FF and intermolecular potentials \cite{AMOEBA, AMOEBA+, CLIFF, MLInteratomicPotentials}.
\begin{equation}
    V^\text{pot, ANA} = V^\text{ele} + V^\text{disp} + V^\text{ind} + V^\text{ct} + V^\text{ex}
\end{equation}
The dispersion interaction ($V^\text{disp}$) is described with dispersion coefficients in conjunction with the Becke-Johnson damping model \cite{BJ, BJXDM}.
Following the model used in AMOEBA/AMOEBA+, induction is included through the Thole damping model ($V^\text{ind}$) and a charge-transfer potential ($V^\text{ct}$) \cite{Thole, AMOEBA, AMOEBA+, Tinker}. 
Exchange and electrostatic interactions ($V^\text{ex}$ and $V^\text{ele}$) are based on anisotropic potentials derived for atomic multipoles following the work by Rackers \textit{et al.}  \cite{AMOEBAChargeDamping, AnisotropicExchange, HIPPO, Tinker}. Atomic multipoles are obtained with our recently introduced equivariant GNN \cite{MultipolesMe}. This model was not further modified for the present work.
A detailed description of the terms is given in the Appendix \ref{sec:appendix_ana}.

\subsubsection{Predicted Parameters}
As in the IPA model, atomic multipoles were obtained using our previously developed equivariant GNN \cite{MultipolesMe}. 
In total, five atomic parameters and five pairwise parameters were predicted by the ANA model. 

Atomic parameters include the atomic polarizability ($\alpha$ in Eq.~(\ref{eq:polarize})), the exponential factor for the electrostatic damping function ($b$ in Eqs.~(\ref{eq:damp_singlesite}) and (\ref{eq:damp_overlap})), the exponential factor used in the damping function of the exchange potential ($b$ in Eqs.~(\ref{eq:damp_exchange}), (\ref{eq:damp_exchange_pair}) and (\ref{eq:damp_B0})), the scaling factor used to adjust the strength of the exchange potential ($k$ in Eq.~(\ref{eq:exchange})), and the number of valence electrons ($q^\text{val}$), which is added to the negative atomic monopole to obtain a scalar that replaces the atomic monopole in the anisotropic exchange potential. 
Further, pairwise $C_6$, $C_8$ and $C_{10}$ parameters were independently predicted for each atom pair as well as the exponents for the short-range induction potential ($A$ and $B$ in Eq.~(\ref{eq:charge_transfer})).

\section{Methods}
\subsection{Differentiable Force Field}
To achieve end-to-end differentiability, all FF terms and parametrization models were implemented in Tensorflow (version 2.5) \cite{TensorflowPaper, TensorflowSoftware}, taking advantage of its automatic differentiation capabilities as well as (batched) GPU accelerated computation.
The particle-mesh-Ewald method (PME) \cite{PME, PME2} implemented in OpenMM (version 7.7) \cite{OpenMM7} was used to obtain the long-range electrostatic contributions for periodic systems.

\subsection{Graph Construction}
The approach applied in this work is based on graphs constructed from the molecular topology, referred to as 'topological graphs'. Topological graphs do not include information about the Euclidean distance between atoms but only atomic connectivity. 
Including geometrical information could be advantageous for certain applications but would require frequent recalculation of parameters, which would limit the performance. In addition, model robustness might suffer from insufficient sampling of intramolecular degrees of freedom. On the other hand, topological graphs may be ill-defined for certain cases and are unable to describe phenomena such as bond forming and breaking. For the envisioned application, i.e., the classical description of molecular motion, the shortcomings of topological graphs are acceptable while presenting a robust and efficient solution. We note that some degree of conformational dependence is present in the overall approach due to the GNN model used for the prediction of atomic multipoles \cite{MultipolesMe}.

No chemical concepts such as bond types or hybridization states were included in the graphs. Hence, graphs only contained a description of the element type of an atom and the presence of a covalent bond between two atoms.
Further information regarding the construction of molecular graphs is given in the Appendix \ref{sec:appendix_graph_construction}.

\subsection{Loss Weighting}
Models were optimized by minimizing the mean square error $L$ between the predicted intermolecular potential energy $V^{\text{pot},\theta}$ and a reference intermolecular potential energy $V^\text{pot,ref}$, which was used as the target property

\begin{equation}\label{eq:dimer_loss}
    L = \frac{1}{N}\sum_i^N w_i(V^\text{pot,ref}(x_i)) \cdot \left[ V^{\text{pot},\theta}(x_i) - V^\text{pot,ref}(x_i) \right] ^2
\end{equation}
with $N$ denoting the batch size, and $i$ iterating over each sample of the batch. 
$w_i$ is a scalar used to weight the contribution of each sample. The importance of each sample was scaled according to its Boltzmann weight in the following manner
\begin{equation}
w_i = \begin{cases}
 1 & \text{if}~r_i > r_\text{eq}\\
 \exp{[-\beta(V^\text{pot,ref}(x_i) - V^\text{pot,ref}(x_\text{eq}))]} & \text{if}~r_i \leq r_\text{eq}
\end{cases}
\end{equation}
Samples beyond the equilibrium distance $r_\text{eq}$ were weighted with $w_i=1$, and samples closer than the equilibrium distance weighted with the Boltzmann weight of the difference between the potential of the given sample and the potential of the equilibrium sample for a given system. The inverse-temperature $\beta$ was used to determine the relative importance. To give more importance to near-equilibrium energy samples towards the end of the training procedure, an exponential decay was used to simulate annealing
\begin{equation}
    T_n = T_0\exp{(-\gamma n)}
\end{equation}
with a decay rate $\gamma$ and an initial temperature $T_0$ and a stopping temperature $T_\text{min}$. 

\subsection{General Training Strategy}
\begin{figure}
    \centering
    \includegraphics[width=0.99\textwidth]{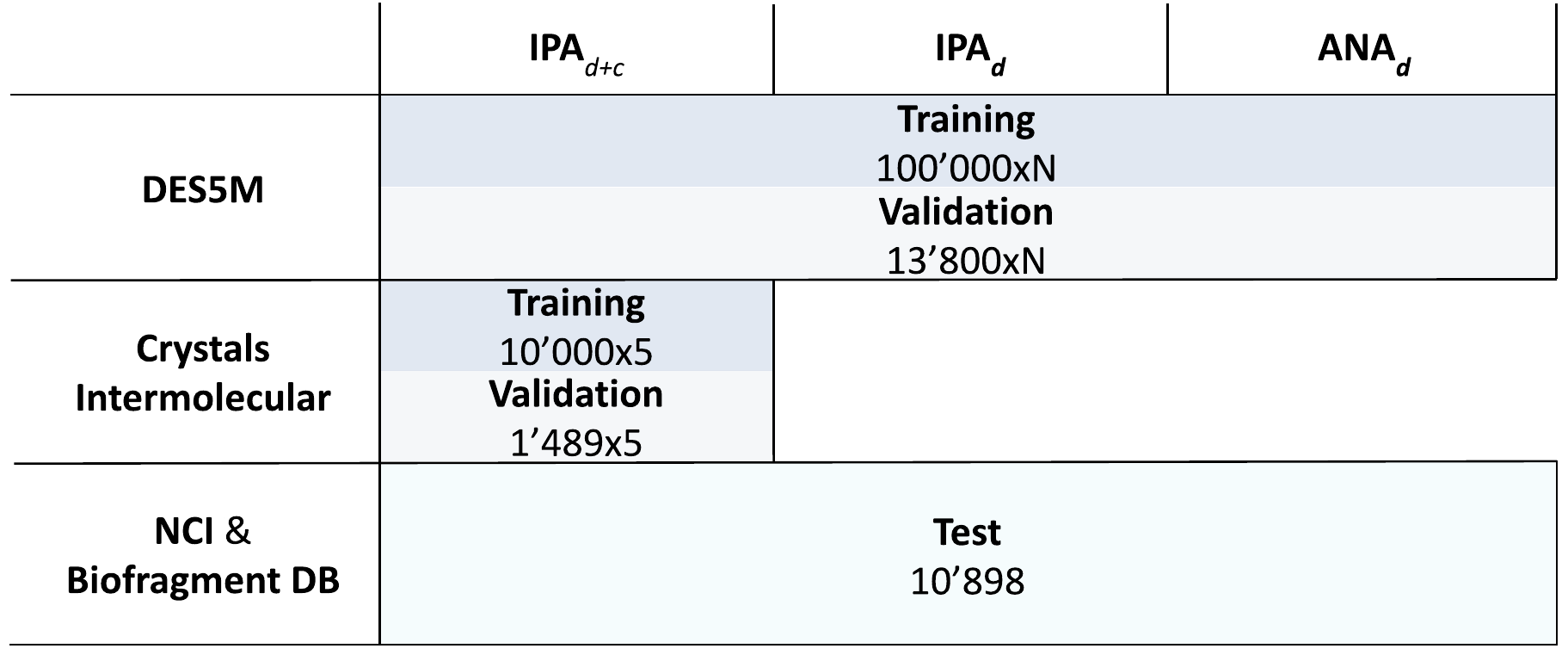}
    \caption{Data sets used for training, validation and testing of the IPA and ANA models. The number of molecular systems in the respective data set is given below the set indication. For dimers in the DES5M data set \cite{DEShawDimers}, a variable number of samples is found, indicated by 'xN'. Details on the data sets are given in Appendix \ref{sec:appendix_trainingsets}.}
    \label{fig:overviewdata sets}
\end{figure}

The recently published DES5M data set \cite{DEShawDimers} (see Appendix \ref{sec:appendix_dimerset}) was used as main data source for the model training (Figure \ref{fig:overviewdata sets}). The data set includes spin-network-scaled MP2 (SNS-MP2) \cite{MP2, SNSMP2, SNS} intermolecular potentials and SAPT0 components \cite{SAPT0, SAPT1} for a large number of small-molecule dimers in vacuum.
In total, $113'800$ dimer sets were included with $100'000$ sets randomly selected for training and the remaining $13'800$ used for validation.
Training was performed over $512$ epochs. During each epoch, a total of $1024$ batches were presented. Each batch contained one dimer set, i.e., all interaction potentials of one dissociation curve or set of clusters for the same two monomer molecules.

\subsection{Training Strategy: IPA Model}
\subsubsection{IPA$_d$ Model}
The IPA model was first optimized against dimer intermolecular potential energies in vacuum from the DES5M data set \cite{DEShawDimers} through minimization of the expression found in Eq.~(\ref{eq:dimer_loss}), with SNS-MP2 energies serving as $V^\text{pot, ref}$. Annealing was performed for the loss weight term $w_i(V^\text{pot, ref}(x_i))$ using the described exponential decay with $T_0=8'000\,$K, $\gamma=-8\cdot 10^{-3}$, and $n$ being incremented after every epoch. Annealing was stopped at $T_{\text{min}}=400\,$K. 
Annealing was of particular importance for the IPA model as the attraction-repulsion potential used is not able to describe very short-ranged interactions accurately. 

The IPA model was regularized with the following term
\begin{equation}
    L_{C9} = \log{(\exp{(C_9 - C_6)} + 1)}^{-1}
\end{equation}
During the training on dimers, an additional term $L_{C6} = C_6^{-1}$ was added.
Both regularization terms were averaged over all interaction pairs of a molecule. 
The model obtained in this manner is called IPA$_{d}$ in the following

\subsubsection{IPA$_{d+c}$ Model}
The IPA$_d$ model was optimized in a second step with respect to intermolecular potentials of molecular crystals. 
Intermolecular potentials were calculated with Quantum Espresso \cite{QE1, QE2, QE3} using the B86bPBE functional in conjunction with the XDM dispersion correction \cite{PBE, PBE1, BJXDM, XDM, XDMB86bPBE}. 
For this step, a total of $11'489$ molecules from the CSD database were used \cite{CSD}. $10'000$ molecules were randomly assigned to the training set and the remaining $1'489$ molecules formed the validation set.
Each molecule entailed five intermolecular potentials, which served as one batch. These five potentials were obtained from the relaxed crystal structure and through the expansion/contraction of the lattice of the relaxed structure. Further details on this data set are given in the Appendix \ref{sec:appendix_crystal_data}.

Again, the mean square error between the predicted and the reference intermolecular potentials was minimized. The electrostatic component was independently calculated with the PME method \cite{PME} and not optimized. Fixed partial charges were obtained as described in Section~\ref{sec:cp_properties}. 
The loss was weighted with
\begin{equation}\label{eq:loss_weighting}
    w_i = \exp{[-\beta(V^\text{pot,ref}(x_i) - V^\text{pot,ref}(x_\text{eq}))/n]}
\end{equation}
Unlike the weighting function used for dissociation curves in Eq.~(\ref{eq:dimer_loss}), the weight was only set to one for the equilibrium structure. In addition, the potential-energy terms were scaled by the number of atoms in the respective molecule. Loss weighting is necessary since contracting/expanding the relaxed crystal structures resulted in highly unfavourable structures in certain cases. 
Training was performed over $512$ epochs. During each epoch, $512$ randomly sampled batches were presented. Each batch contained five intermolecular potentials of one specific molecule.
Throughout, the temperature was set to $T_0=128\,$K. A cutoff of $10\,$\AA\, was used for nonbonded interactions.

\subsection{Training Strategy: ANA Model}
Optimization of the ANA$_d$ model followed the same procedure used to optimize the IPA$_{d}$ model based on the dimer data set, except for the following differences.
Since the functional form of the ANA model allows for a more accurate description of short-range interactions, the annealing schedule was modified to $T_0=40'000\,$K, $\gamma=-7.5\cdot 10^{-3}$, and $T_\text{min}=2'000\,$K.
Furthermore, the $L_{C_6}$ and $L_{C_9}$ terms were replaced with additional loss terms for the energy components as the ANA model permits a decomposition of the total energy into components, which can be related to SAPT components \cite{SAPT0, SAPT1}. 
\begin{equation}
    L_{\text{SAPT}} = \kappa \sum_\lambda\frac{1}{N}\sum_i^N w_i(V^\text{pot,ref}(x_i)) [V_\lambda^{\text{pot},\theta}(x_i) - V_\lambda^\text{pot, SAPT}(x_i)]^2
\end{equation}
with $\lambda$ iterating over all SAPT components, i.e., exchange, induction, electrostatic and dispersion.
The weights $w_i$ were calculated for the total reference potential.

The SAPT loss term was scaled by a factor $\kappa$ and added to the total loss in Eq.~(\ref{eq:dimer_loss}). The weight for the total energy component with respect to the SNS-MP2 calculation was kept at $1$.
$\kappa$ was initially set to $0.5$ and reduced to $0.01$ after $256$ epochs had passed. 
We note that the inclusion of the SAPT loss term is a double-edged sword. Preliminary investigations indicated that the SAPT loss term serves on one hand as a regularizer, which also accelerates convergence. On the other hand, larger values for $\kappa$ limit the degree of error cancellation between potential-energy terms. It is further important to keep in mind that the accuracy of the employed SAPT0 method is lower than SNS-MP2. Specifically, for the considered subset of the DES5M data set, a MAE of $4.19\,$kJ/mol and $0.77\,$kJ/mol was found when using the weighting in Eq.~(\ref{eq:loss_weighting}) with T$=2000\,$K.
With a mean error of $-4.11\,$kJ/mol and $-0.71\,$kJ/mol, SAPT0 overbinds relative to the SNS-MP2 results. This observation is consistent with previous benchmarks \cite{SAPT1}. 

\subsection{Calculation of Condensed-Phase Properties}\label{sec:cp_properties}
The IPA$_{d+c}$ model was also applied to calculate properties of condensed-phase systems.
The simulations were performed under periodic boundary conditions using OpenMM \cite{OpenMM7}. 
Up to $32$ conformers were generated for each molecules with the ETKDG conformation generator \cite{ETKDG} as implemented in the RDKit \cite{RDKIT}. An RMS pruning threshold of $0.1\,$\AA\, was used. 
Monopoles were predicted for each conformation using our previous equivariant GNN \cite{MultipolesMe}, and then averaged over the conformational ensemble to yield the fixed partial charges for the simulation.
The same partial charges were used for the simulations in the condensed phase and in the gas phase.
Bonded terms were parametrized with OpenFF 2.0 \cite{SageFF} since the IPA model handles only intermolecular interactions. Further details regarding simulation setups are given in the Appendix \ref{sec:appendix_simulation_protocol}.

\subsubsection{Lattice Energy}
In the present work, the lattice energy $V^\text{lattice}$ is approximated as
\begin{equation}
    V^\text{lattice} \approx \frac{V^\text{pot, inter}}{Z}
\end{equation}
ignoring the contribution of the intramolecular interactions. $Z$ refers to the number of molecules in the unit cell, and $V^\text{pot, inter}$ to the total intermolecular potential energy for a unit cell with $Z$ molecules under periodic boundary conditions. $V^\text{pot, inter}$ was calculated for the experimental geometries without relaxation. A list of CSD codes is given as Supporting Information.

\subsubsection{Heat of Vaporization}
The heat of vaporization was computed from the difference between the mean potential energy in the gas phase $\langle V^\text{pot, gas} \rangle$ and the mean potential energy per molecule in the condensed phase $\langle V^\text{pot, liq} \rangle$ corrected by a factor of RT,
\begin{equation}
    H_\text{vap} = \langle V^\text {pot, gas} \rangle - \langle V^\text{pot, liq} \rangle + RT 
\end{equation}
where $R$ is the gas constant, and $T$ the absolute temperature.

\subsubsection{Density}
The condensed-phase density was calculated as the total mass $m_\text{box}$ in the simulation box divided by its average volume $\langle V_\text{box} \rangle$
\begin{equation}
    \rho = \frac{m_\text{box}}{\langle V_\text{box} \rangle}.
\end{equation}

\subsubsection{Static Dielectric Constant}
Static dielectric constants $\epsilon$ were obtained from the fluctuation of the dipole $M$ of the system as described in Ref.~\cite{CPEquations},
\begin{equation}
    \epsilon = 1 + \frac{4\pi}{3k_B T\langle V_\text{box} \rangle}(\langle M^2\rangle - \langle M\rangle^2)
\end{equation}

\subsubsection{Isothermal Compressibility}
Similarly, isothermal compressibilities $\kappa$ were obtained for fluctuations of the system volume $V_\text{box}$ \cite{CPEquations},
\begin{equation}
    \kappa = -\frac{1}{V_\text{box}}\bigg(\frac{\partial V_\text{box}}{\partial P}\bigg)_{N, T} \approx \frac{\langle V_\text{box}^2 \rangle - \langle V_\text{box} \rangle^2}{k_B T\langle V_\text{box} \rangle}
\end{equation}
where $P$ is the system pressure.

\subsubsection{Thermal Expansion Coefficient}
Thermal expansion coefficients $\alpha$ were computed via the following relation \cite{CPEquations},
\begin{equation}
    \alpha = \frac{1}{V_\text{box}}\bigg(\frac{\partial V_\text{box}}{\partial T}\bigg)_{N, P} \approx \frac{\langle V_\text{box}H_l \rangle - \langle V_\text{box} \rangle \langle H_l \rangle}{k_B T^2\langle V_\text{box} \rangle} ,
 \end{equation}
where $H_l$ is the total enthalpy of the box.

\section{Results and Discussion}
The performance of the IPA$_d$, IPA$_{d+c}$ and ANA$_d$ models to describe intermolecular interactions were investigated for a wide range of systems and environments. For the IPA$_{d+c}$ model, emphasis is put on the performance in condensed-phase systems, which are generally challenging for non-classical models. With its more sophisticated functional form, the ANA model is applied to intermolecular potentials of small-molecule dimers, which permits a direction comparison with first-principle and DFT methods.
Finally, further explorations of parameters learned by both approaches are presented.
We note that the ANA model has so far not been parametrized and applied to condensed-phase systems. 

\subsection{Intermolecular Potentials in Vacuum}
Small-molecule dimers are the largest systems that can be treated with highly accurate wave-function methods. As such they present a valuable validation case to probe the accuracy of the description of specific interactions.
For this purpose, several established non-covalent interaction benchmarks were taken as test sets from the Biofragment database \cite{BioFragmentDB, NBC10_1, NBC10_2, HSG, UBQ, JSCH, ACHC} and the non-covalent interaction (NCI) atlas \cite{S66x8, NCIDispersion, NCIRepulsion, NCIHB300SPX×10, NCIHB375×10} (Figure \ref{fig:overviewdata sets}). In addition, the models are tested on the supramolecular S12L \cite{S12L} data set. For systems in the S12L, binding energies calculated with quantum diffusion Monte Carlo were used as reference \cite{S12LT}.
In all cases, only systems consisting of neutral monomers with more than two atoms were included.
The performance results on the benchmarking sets with a total of $10'894$ unique data points are summarized in Table \ref{tab:dimer_results}. The full error statistics is given in Table S1 - S3 in the Supporting Information.

\begin{table}[htb]
    \centering
    \begin{tabular}{l l c | c c c}
    \hline
    & & & \multicolumn{3}{c}{MAE for interactions in vacuum [kJ/mol]}\\
    Source & Data set & Data points & IPA$_{d+c}$ & IPA$_{d}$ & ANA$_d$ \\\hline
    NCI & S66x8 \cite{S66x8} & 528 & 2.9 & 2.3 & 1.1\\
    Biofrag. & SSI \cite{BioFragmentDB} & 2596 & 1.4 & 1.1 & 1.0\\
    Biofrag. & BBI \cite{BioFragmentDB} & 100 & 3.3 & 0.8 & 1.1\\
    Biofrag. & UBQ \cite{UBQ} & 81 & 3.9 & 1.7 & 1.4\\
    Biofrag. & ACHC \cite{ACHC} & 54 & 2.7 & 5.8 & 2.7\\
    Biofrag. & JSCH \cite{JSCH} & 123 & 5.9 & 6.5 & 2.3\\
    Biofrag. & HSG \cite{HSG} & 16 & 1.8 &  1.3 & 0.9\\
    NCI & D1200 \cite{NCIDispersion} & 401 & 4.2 & 2.1 & 1.9\\
    NCI & D442x10 \cite{NCIDispersion} & 1230 & 4.6 & 3.1 & 2.9\\
    NCI & R739x5 \cite{NCIRepulsion} & 1370 & 4.5 & 3.7 & 3.6\\
    NCI & HB375x10 \cite{NCIHB375×10} & 3750 & 4.4 & 4.9 & 1.9\\
    NCI & HB300SPXx10 \cite{NCIHB300SPX×10} & 640 & 3.5 & 3.8 & 2.3\\
    - & S12L \cite{S12L, S12LT} & 5 & 26.5 & 58.4 & 64.5\\
    \end{tabular}
    \caption{Mean absolute error (MAE) of the IPA$_d$, IPA$_{d+c}$ and ANA$_d$ models on the benchmarking sets in vacuum. IPA$_{d+c}$ refers to the fixed-charge model, which included training on crystal structures. The IPA$_{d}$ and the polarizable ANA$_d$ models were trained exclusively on dimer interaction potentials from the DES5M data set \cite{DEShawDimers}. The full error statistic is given in Tables S1 - S3 in the Supporting Information.}
    \label{tab:dimer_results}
\end{table}

The results of the three models investigated in this study are shown in Table \ref{tab:dimer_results}.
Consistent results over these diverse sets are observed. For comparison, reference values are given for the data sets in the Biofragment database where available (Table \ref{tab:reference_values}). Three methods were chosen to represent classical force fields (CHARMM General FF (CGenFF) \cite{CGENFF}), semi-empirical models (PM6-DH2 \cite{PM6DH2}), and DFT methods (PBE0-D3 \cite{PBE0, D3, D3BJ}). The IPA models perform comparable to previously reported results for empirical (FF) or semi-empirical models. In general, the simpler models are outperformed by the ANA$_d$ model, which achieves for some data sets an accuracy comparable to dispersion corrected hybrid functionals like PBE0-D3BJ. Exceptions are the BBI data sets, where the IPA$_d$ model performs better than ANA$_d$, and the S12L data set for which IPA$_{d+c}$ outperforms the other models.

\begin{table}[htb]
    \centering
    \begin{tabular}{l c | c c c}
    \hline
    & & \multicolumn{3}{c}{MAE for interactions in vacuum [kJ/mol]}\\
    Data set & Data points & CGenFF & PM6-DH2 & PBE0-D3 \\\hline
    S66x8 \cite{S66x8} & 528 & - & 3.3 \cite{S66x8PM6} & 1.0 \cite{S66x8REF} \\
    SSI \cite{BioFragmentDB} & 2594 & 1.3 & 1.1 & 0.5\\
    BBI \cite{BioFragmentDB} & 100 & 2.1 & 2.9 & 0.3\\
    UBQ \cite{UBQ} & 80 & 3.7 & 1.4 & -\\
    ACHC \cite{ACHC} & 54 & -  & - & 1.8\\
    HSG \cite{HSG} & 16 & 1.3 & 1.8 & 1.3 \\
    \end{tabular}
    \caption{Mean absolute error (MAE) for data sets in the Biofragment database and S66x8 for the classical force field CGenFF \cite{CGENFF}, the semi-empirical model PM6-DH2 \cite{PM6DH2}, and the DFT method PBE0-D3 \cite{PBE0, D3, D3BJ}. Reference values were taken from the publication of the data set if not indicated otherwise. Values were converted to kJ/mol using a factor of $4.184$. Note that PBE0-D3 does not use the same basis sets in all cases. Values for the largest available basis set (def2-QZVP or aug-cc-pVTZ) and counterpoise correction were chosen if available.}
    \label{tab:reference_values}
\end{table}

Comparison between IPA$_{d+c}$ and IPA$_{d}$ suggests that additional training on crystal data points does not necessarily lead to a strong negative impact on the description accuracy of dimers in vacuum. This is surprising for two reasons: (i) The method used to compute the potential energy of crystal structures is considerably less accurate than the methods used to compute the dimer interaction potentials in the benchmarking data sets and the DES5M training data set. (ii) Polarization effects, for which the IPA models must implicitly account, differ between vacuum and condensed-phase environments. This effect is visible in the case of the S12L data set, where IPA$_{d+c}$ clearly outperforms the IPA$_d$ model.
Interestingly, training on crystal potential energies appears to provide a regularizing effect for the IPA$_{d+c}$ model, offering not the most accurate but the most consistent results over a wide range of applications beyond dimers (see below).

The results for the five neutral systems of the revisited S12L data set are worthy of special attention. For these supramolecular systems, a consistent overestimation of the interaction potential is observed for the ANA$_d$ model with a mean error equal to the mean absolute error (Table S1 in the Supporting Information).
A large contribution might be due to the missing treatment of many-body effects, which was noted by the creator of S12L \cite{S12L} and further discussed for the revisited values \cite{S12LT}. Estimation of the three-body dispersion contribution would explain approximately half of the overbinding reported here \cite{S12L}. 
Previous work observed similar effects with ML models \cite{PhysicsBasedML}. However, despite using a dispersion correction with many-body effects, Ref.~\cite{PhysicsBasedML} reported comparable results to our models. The performance could only be improved through the inclusion of the same structures in the training set \cite{PhysicsBasedML}.
In a similar vein, a false balance between many-body effects (e.g., between the potential energy due to induced dipole and pairwise interactions such as dispersion and the short-range induction potential) might contribute further to this effect. A weak overestimation of pairwise interactions might not be noticeable for small systems but could be amplified for larger systems. Overbinding of SAPT and its components could further contribute to this observation. 
Thus, the S12L data set may be exposing a possible limitation of polarizable models and small molecule data sets.
In this context, we also note a recent study where the authors observed considerable discrepancies between reference methods for large complexes \cite{QCMRef}.
Investigating which effects need to be included to permit the generalization to larger systems and the condensed phase is therefore an important open question. 
Developing additional benchmarking data sets with a broader coverage of the space from medium to large structures could be highly beneficial to validate such efforts.

\subsubsection{Model Parameters}
The influence of feature size and the number of GNN layers was investigated for the IPA$_{d+c}$ model. 
For this, a range of models with varying feature dimensionality and GNN layers was trained on the DES5M data set as described in the Methods section. 
Shorter epochs were used, i.e., $512$ batches instead of $1'024$, while all other parameters remained unchanged.
The performance was assessed on the basis of absolute weighted (T$=400\,$K) errors for the DES5M validation set, which contains 150 molecules resulting in $13'800$ dimer sets and a total of $471'149$ data points. 

\begin{figure}[htb]
    \centering
    \includegraphics[width=0.99\textwidth]{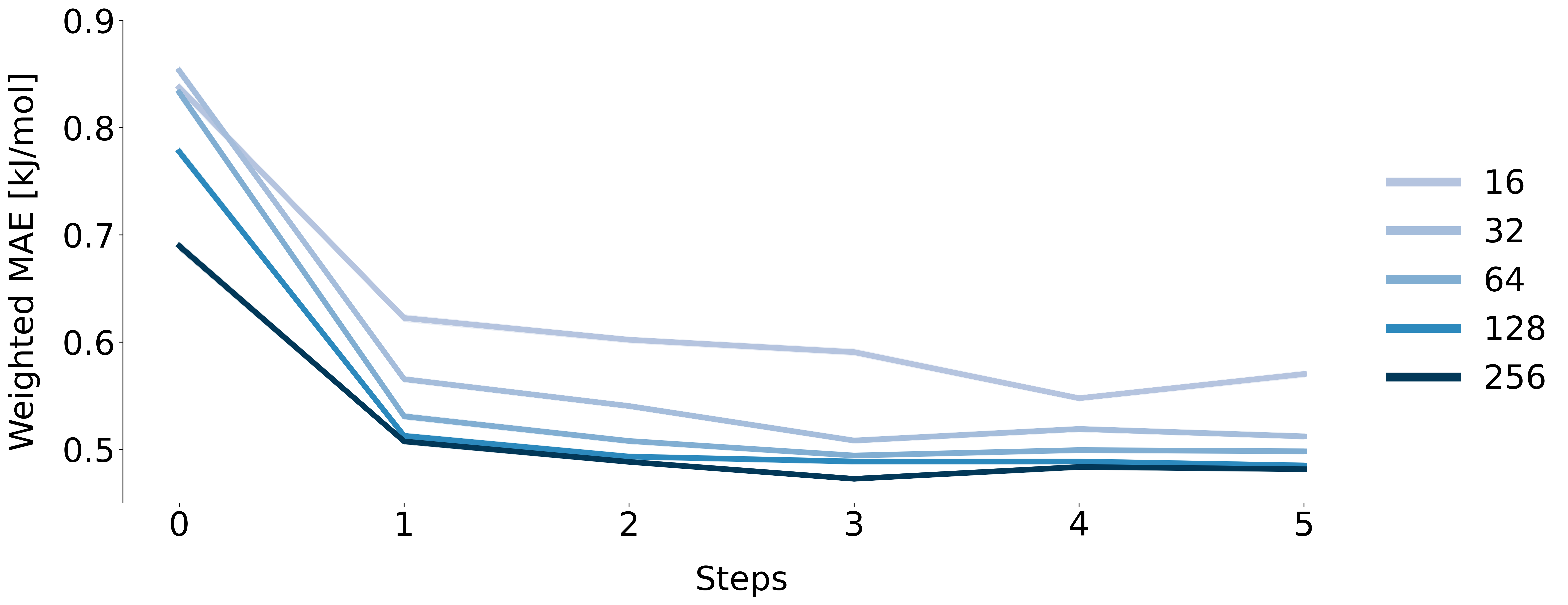}
    \caption{Performance of the IPA$_{d+c}$ model on the DES5M validation set \cite{DEShawDimers} for a given numbers of graph-update steps and feature dimensions (16 - 256). The validation set contains 150 molecules giving $13'800$ dimer sets and a total of $471'149$ data points.}
    \label{fig:size_validation}
\end{figure}

The results shown in Figure \ref{fig:size_validation} support the inclusion of information about the atomic neighbourhood. The first two coordination shells are particularly beneficial, with marginal benefits or even detrimental effects for larger numbers of steps. 
Purely local models (number of steps = 0), which only include information about the atomic element, perform considerably worse than non-local models. Nevertheless, even for this most simple case, acceptable performance is observed, which might be due to the monopoles already capturing the most important features of atomic interactions.
Interestingly, increasing the numbers of parameters (i.e., feature dimensions) of the model itself does not seem to provide benefits beyond a given range.
It is important to note that the molecular graphs for the considered DES5M validation set contain comparatively small monomers with a mean graph diameter of $4.2$ and a maximum graph diameter of $8$. Some of these conclusions might change for larger molecules.
It could be particularly illustrative to investigate the potential benefit of additional graph-updating steps for large aromatic systems with non-local effects due to substituents such as nitro groups or long and branched systems.

\subsection{IPA$_{d+c}$ Model -- Condensed-Phase Properties}
Reproducing properties of condensed-phase systems is an important validation task for intermolecular potentials. QM methods are in general not feasible to simulate such systems, and the wide range of interactions poses considerable challenges to ML models.
In the following, the IPA$_{d+c}$ model is validated on various condensed-phase properties of pure organic liquids.

\subsubsection{Intermolecular Potentials in Crystals}
Intermolecular potential energies calculated for a wide range of molecular crystals were used to parametrize the IPA$_{d+c}$ model in addition to the dimers in vacuum. 
Table \ref{tab:intermolecular_ipa} shows the mean absolute error on the molecular crystals for the IPA$_{d+c}$ and the IPA$_d$ models.
Only equilibrium structures of the crystals were included. The training set and test set contained $10'000$ and $1'507$ data points, respectively.
As can be expected, the IPA$_d$ model performs considerably worse than the IPA$_{d+c}$ model. As the latter model performs also comparatively well on dimers in vacuum, these results point at a general advantage for the training of IPA-type models on both gas-phase and condensed-phase data.
For this reason, only the IPA$_{d+c}$ model is used in the following.

\begin{table}[htb]
    \centering
    \begin{tabular}{l | c c}
    \hline
    & \multicolumn{2}{c}{MAE for intermolecular potentials in crystals [kJ/mol]}\\
    Data set & IPA$_{d+c}$  & IPA$_d$ \\\hline
    Training & 5.3 & 28.7\\
    Test & 5.3 & 28.2\\
    \end{tabular}
    \caption{Mean absolute errors (MAE) for intermolecular potentials of molecular crystals at equilibrium calculated with the IPA$_{d+c}$ and the IPA$_d$ models. The training set and test set contained $10'000$ and $1'507$ data points, respectively.}
    \label{tab:intermolecular_ipa}
\end{table}

\begin{figure}[htb]
    \centering
    \includegraphics[width=0.99\textwidth]{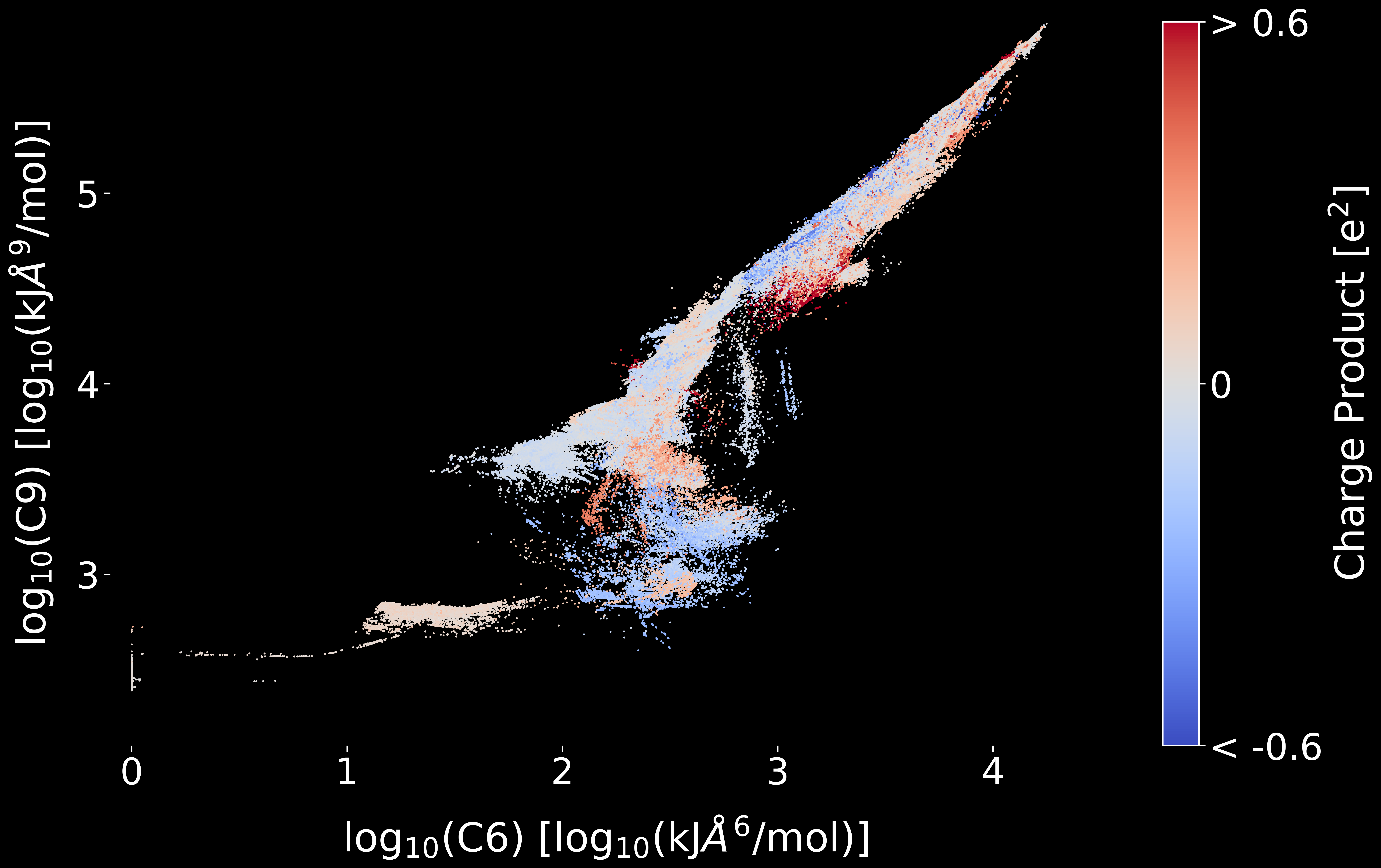}
    \caption{Predicted C6 and C9 parameters by the IPA$_{d+c}$ model on log scale for all atom pairs in the data set of intermolecular potentials of crystals (11'507 molecules). The color indicates the strength and sign of the electrostatic interaction (blue = attractive, red = repulsive).}
    \label{fig:parameter_landscape}
\end{figure}

To obtain a picture of the range of van der Waals parameters predicted by the model, Figure \ref{fig:parameter_landscape} shows the C$_6$ and C$_9$ parameters from the IPA$_{d+c}$ model for all pairwise interactions in the crystal data set. Further information is given in Figures S1 and S2 in the Supporting Information, showing the resulting well-depth and the minimum distance by atom pairs.
The nature of the electrostatic interaction is indicated by the color, with blue for attractive interactions and red for repulsive interactions. 
Interestingly, a large part of the correlation between C$_6$ and C$_9$ parameters might be captured by a power law. 
This observation could potentially be used to construct FF based on a single parameter and appropriate scaling laws, for instance based on a notion of atomic volumes. 
Furthermore, the predicted C$_6$ and C$_9$ parameters form a continuum over large ranges. While there are some distinct islands, in particular interactions with hydrogens in the bottom left, the results in Figure \ref{fig:parameter_landscape} are nevertheless an indication that the model takes advantage of continuous atom types.

\subsubsection{Lattice Enthalpies: X23 and G60} 
Performance on crystal structures was evaluated on the X23 benchmark \cite{X23Johnson, X23Tkatchenko} using the revised values found in Ref.~\cite{X23Revised} as well as the G60 data set \cite{G60} (Table \ref{tab:x23}).
Two settings were considered: For setting (1), labelled with 'Non-relaxed', the lattice energy was approximated with the intermolecular potential for the non-relaxed experimental geometries. 
In setting (2), labelled with 'Relaxed + intra', structures were relaxed and the potential-energy difference between the gas-phase minimum conformation and the crystalline phase minimum was included. For the second setting, six systems had to be excluded due to problems with the bonded terms from OpenFF (i.e., X23: CO$_2$ and UREAXX12; G60: CTMTNA03, METNAM08, MTNANL, OCHTET13).

For both data sets, the cohesive energy is systematically underestimated with the IPA$_{d+c}$ model (i.e., positive mean errors in Table \ref{tab:x23}) and the MAE is above chemical accuracy. 
Nevertheless, the IPA$_{d+c}$ model reproduces lattice energies more accurately than most existing models reported so far in the literature, e.g. DFTB-D3 with a MAE of $10.38\,$kJ/mol for the X23 data set \cite{X23DFTBD3}, except for some of the best performing dispersion corrected DFT functionals such as PBE0-MBD with a MAE of $3.9\,$ kJ/mol on the X23 data set \cite{X23Tkatchenko, X23Johnson}. 
This shows that the accurate description of the lattice energy of molecular crystals remains a particularly challenging problem, which will continue to serve as an important reality check, specifically for models that are not parametrized on condensed-phase systems. 

\begin{table}[htb]
    \centering
    \begin{tabular}{l | c c | c c}
    \hline
    & \multicolumn{2}{c|}{MAE for lattice enthalpies [kJ/mol]} & \multicolumn{2}{c}{Mean error for lattice enthalpies [kJ/mol]}\\
    Data set & Non-relaxed & Relaxed + intra & Non-relaxed & Relaxed + intra\\\hline%
    X23 & 7.1 & 6.1 & 4.2 & 2.2 \\
    G60 & 9.5 & 12.1 & 5.4 & 9.5 \\
    \end{tabular}
    \caption{Mean absolute error (MAE) and mean error of the IPA$_{d+c}$ model for the X23 and G60 benchmark sets based on revised values provided by Ref.~\cite{X23Revised} and reference values collected in Ref.~\cite{G60}. The full error statistics are provided in Table S4 and S5 in the Supporting Information.}
    \label{tab:x23}
\end{table}

\subsubsection{Pure Liquid Properties}
Properties of pure organic liquids are commonly used to validate classical FF, serving in many cases also as parametrization targets (e.g., \cite{OPLS, 2016H66, OpenFFCP}).
To further explore the performance of the IPA$_{d+c}$ model beyond vacuum and the crystalline phase, applications to the liquid phase are shown in the following.
Three benchmarks covering a wide range of systems and properties are considered for this:
(i) $13$ sulfur compounds taken from the publication of the OPLS4 release \cite{OPLS4}, (ii) $29$ molecules containing H, C, O taken from a recent investigation of condensed-phase parametrization of OpenFF \cite{OpenFFCP}, and (iii) $57$ organic compounds from the GROMOS 2016H66 validation \cite{2016H66}.
Unlike the referenced FF, parametrization of the IPA$_{d+c}$ model did not include experimental liquid properties such as the density or the heat of vaporization. Therefore, pure liquid properties present an interesting test case for this model.
For the liquid simulations with the IPA$_{d+c}$ model, bonded interactions (i.e., bonds, angles, dihedrals, and 1-4 nonbonded interactions) were treated with OpenFF 2.0 \cite{SageFF}, while all other interactions (i.e., nonbonded terms) were treated with the IPA$_{d+c}$ model.
A detailed description of the simulation protocol is given in the Appendix \ref{sec:appendix_simulation_protocol}. \\

\textbf{Sulfur Compounds}.
Systems containing sulfur are challenging for fixed-charge FF due to the polarizability of sulfur and the presence of higher-order multipole components. Recent work on the OPLS4 FF \cite{OPLS4} improved the performance on several challenging motifs, including sulfur interactions and $\sigma$-holes. Specifically, OPLS4 improved the RMSE of the heat of vaporization ($H_\text{vap}$) for 13 sulfur-containing systems by more than $1\,$kJ/mol (i.e., 0.3~kcal/mol) compared to the previous OPLS3 version \cite{OPLS4}.
With an RMSE of $2.5$\, kJ/mol for $H_\text{vap}$, the IPA$_{d+c}$ model performs comparable to OPLS4 (Table \ref{tab:hvap_sulfur}).
This result is remarkable for several reasons. First, unlike the IPA model, OPLS4 uses virtual-sites to represent lone pairs and anisotropic Lennard-Jones interactions, which were specifically introduced to improve the description of systems containing sulfur and halogens. Second, the OPLS FF family was specifically parametrized with respect to liquid properties such as $H_\text{vap}$ \cite{OPLS}, whereas the IPA$_{d+c}$ model was only trained on QM potential energies. Third, the intramolecular potential of OPLS4 is jointly optimized with the intermolecular potential, allowing for a larger degree of consistency between the two parts. 
It is likely that considering the above points in future work on the IPA model could result in further improvements.

\begin{table}[htb]
    \centering
    \begin{tabular}{l | c c}
    \hline
    & \multicolumn{2}{c}{RMSE for pure liquid properties of sulfur compounds}\\
    Property & IPA$_{d+c}$ & OPLS4 \cite{OPLS4} \\\hline
    $H_\text{vap}$ [kJ/mol]& 2.5 & 2.6\\
    $\rho$\, [kg$\cdot$m$^{-3}$] & 26.0 & -\\
    \end{tabular}
    \caption{Root-mean-square error (RMSE) for the heat of vaporization ($H_\text{vap}$) and density ($\rho$) for 13 sulfur-containing compounds investigated in Ref.~\cite{OPLS4}. 
    Values for OPLS4 from Ref.~\cite{OPLS4} are given as comparison and were converted from kcal/mol to kJ/mol by a factor of $4.184$. Note that densities were not reported for OPLS4 in Ref.~\cite{OPLS4}. The full error statistics are provided in Table S6 in the Supporting Information. The individual numerical values are given in Table S11.}
    \label{tab:hvap_sulfur}
\end{table}

\textbf{Test systems from OpenFF}.
To gain a better understanding for the role of the bonded terms taken from OpenFF, results for 29 pure liquids from a recent benchmark of OpenFF are presented here \cite{OpenFFCP}. The compounds contain only H, C, and O. The referenced work is particularly interesting for its investigation of opposing forces during the parametrization with respect to mixing enthalpies, vaporization enthalpies, and densities. 
Only pure liquid properties were considered here.

As can be seen in Table \ref{tab:openff_properties}, similar errors are observed for the IPA$_{d+c}$ model and the standard OpenFF 1.0. Re-optimization of the FF with respect to the pure liquid properties of the training set (also compounds containing only H, C, and O) improved the accuracy of OpenFF 1.0 considerably \cite{OpenFFCP} (Table \ref{tab:openff_properties}). The authors observed thereby opposing gradient components for the simultaneous optimization with respect to densities and heats of vaporization. No further liquid properties were considered in Ref.~\cite{OpenFFCP} (such as dielectric permittivity, thermal expansion coefficient, etc.). It would thus be interesting to see the performance of the re-optimized OpenFF (termed 'Pure only') on other properties.

The fact that all models shown in Table~\ref{tab:openff_properties} use the same functional form with very similar bonded terms may indicate that liquid properties cannot be reproduced more accurately without either improving the description of the bonded interactions and/or extending the functional form, for instance through the use of a polarizable model and multipoles.
We note that the simulations with the IPA$_{d+c}$ model used bonded terms from OpenFF 2.0 were employed, while OpenFF 1.0 was employed in Ref.~\cite{OpenFFCP}.

\begin{table}[htb]
    \centering
    \begin{tabular}{l | c c c}
    \hline
    & \multicolumn{3}{c}{RMSE for pure liquid properties of the OpenFF compounds}\\
    Property & IPA$_{d+c}$ & OpenFF 1.0 \cite{OpenFFCP} & OpenFF ('Pure only') \cite{OpenFFCP} \\\hline
    $H_\text{vap}$\, [kJ/mol]& 9.2 & 9.9 & 7.5\\
    $\rho$\, [kg$\cdot$m$^{-3}$] & 32.1 & 30.0 & 18.0\\
    \end{tabular}
    \caption{Root-mean-square error (RMSE) for pure liquid properties of 29 systems used as test set in Ref.~\cite{OpenFFCP}. Values for OpenFF were taken from the referenced publication. The label 'Pure only' refers to a version of OpenFF 1.0, which was re-optimized on the densities and heats of vaporization of the training set in Ref.~\cite{OpenFFCP}. The full error statistics are provided in Table S7 in the Supporting Information. The individual numerical values are given in Table S12.}
    \label{tab:openff_properties}
\end{table}

\textbf{Test systems from GROMOS 2016H66}.
The 57 pure liquids from the GROMOS 2016H66 \cite{2016H66} release include extended coverage of the chemical space and additional properties such as
the isobaric thermal expansion coefficient ($\alpha$), the static relative dielectric permittivity ($\epsilon$), and the isothermal compressibility ($\kappa$).
For the considered properties, the IPA$_{d+c}$ model performs comparable to the 2016H66 FF (Table \ref{tab:gromos_properties}). While $H_\text{vap}$ is less accurately reproduced by the IPA$_{d+c}$ model, smaller errors are observed for the remaining properties.
Note that 2016H66 was parametrized on the $H_\text{vap}$ and density values of $27$ of the considered $57$ molecules.
The observation that the IPA$_{d+c}$ model outperforms 2016H66 on properties that were not used for its parametrization may be an indication that the 'bottom-up' approach of the IPA approach, focusing on the reproduction of the PES, is a valuable parametrization strategy. It further demonstrates that there is still room for improvement for the fixed-charge FF model. 

\begin{table}[htb]
    \centering
    \begin{tabular}{l | c c}
    \hline
    & \multicolumn{2}{c}{RMSE for pure liquid properties of the 2016H66 compounds}\\
    Property & IPA$_{d+c}$ & GROMOS 2016H66 \cite{2016H66}\\\hline
    H$_\text{vap}$\, [kJ/mol]& 4.5 & 3.5\\
    $\rho$\, [kg$\cdot$m$^{-3}$] & 26.3 & 32.4\\
    $\alpha$\, [$10^{-4}\, \text{K}^{-1}$] & 1.7 & 4.4\\
    $\epsilon$\, [$1$] & 12.8 & 14.0\\
    $\kappa$\, [$10^{-5}\, \text{bar}^{-1}$] & 1.8 & 3.6\\ 
    \end{tabular}
    \caption{Root-mean-square error (RMSE) for pure liquid properties of 57 systems used in the calibration and validation of the GROMOS 2016H66 FF \cite{2016H66}. Values for GROMOS 2016H66 were taken from the referenced publication. The full error statistics are provided in Table S8 in the Supporting Information. The individual numerical values are given in Table S13.}
    \label{tab:gromos_properties}
\end{table}

\subsection{ANA Model}
\subsubsection{Learned Parameters}
As the ANA approach cannot be used yet for condensed-phase simulations, we validated the ANA$_d$ model by comparing the predicted molecular polarizabilities and intermolecular $C_6$ dispersion coefficients to experiment.
The data set consists of molecular polarizability values for $87$ compounds \cite{Alphas1, Alphas2, Alphas3, Alphas4, Alphas5, Alphas6} and $C_6$ coefficients for $231$ molecular pairs \cite{C6Parameters}. 
Since the ANA model predicts atomic parameters, the molecular polarizability is obtained as the sum of all atomic contributions. Dispersion coefficients are summed over all intermolecular atom pairs as in previous work \cite{TSDispersion, D3, XDM}.

\begin{figure}[htb]
    \centering
    \includegraphics[width=0.99\textwidth]{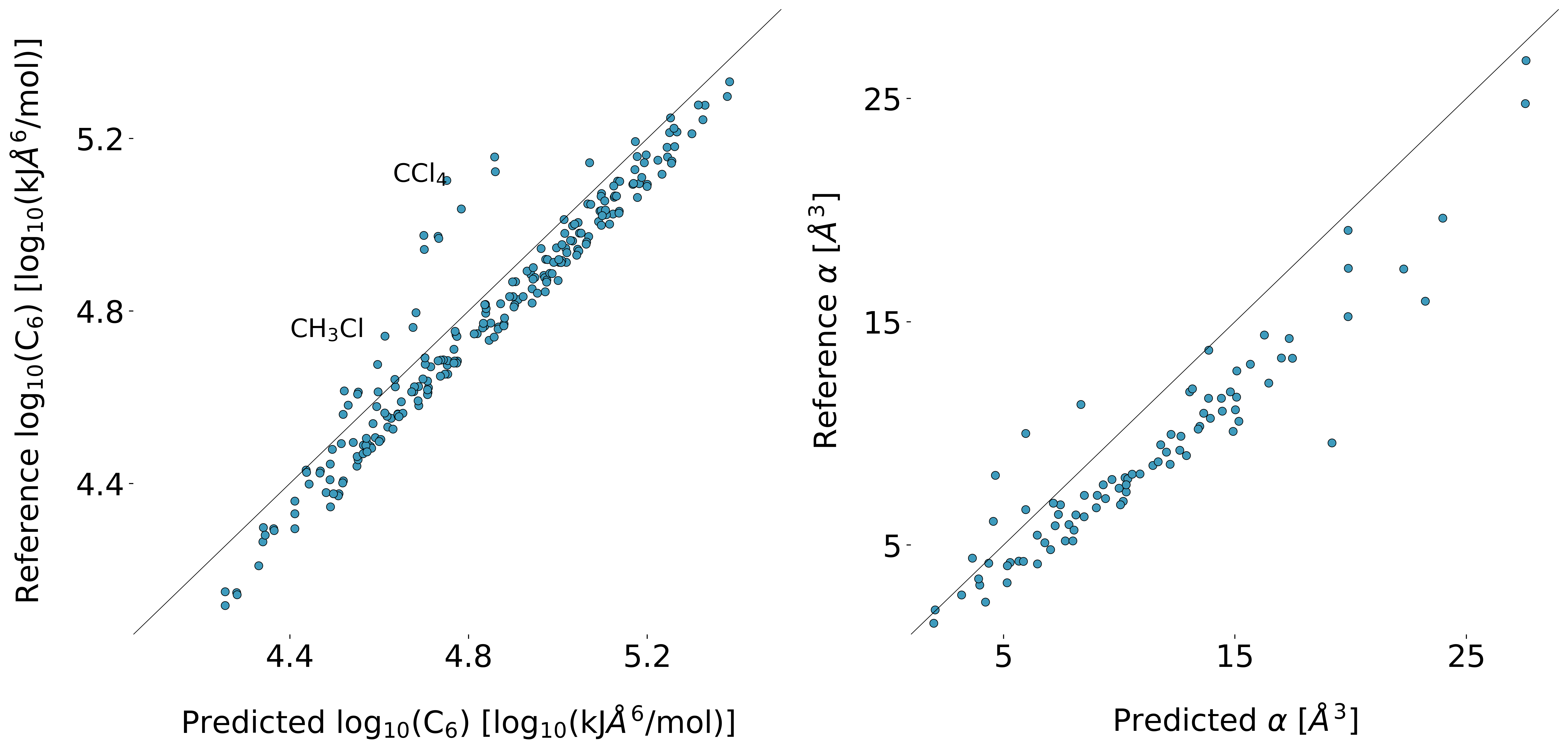}
    \caption{Comparison of the $C_6$ dispersion coefficients (right) and molecular polarizabilities (left) obtained from experiment and predicted by the ANA model. Experimental values for the molecular polarizability of 87 compounds were taken from Refs.~\cite{Alphas1, Alphas2, Alphas3, Alphas4, Alphas5, Alphas6}, and experimental $C_6$ coefficients of $231$ molecular pairs were taken from Ref.~\cite{C6Parameters}.}
    \label{fig:ana_parameters}
\end{figure}

Figure \ref{fig:ana_parameters} shows the comparison between predicted and experimental values. Mean absolute relative errors of $19.4\,$\% and $2.1\,$\% and Spearman correlation coefficients of $0.95$ and $0.94$ are observed for the $C_6$ coefficients and the molecular polarizabilities, respectively. 
For the $C_6$ coefficients, two sets of outliers are found. One set includes interactions with tetrachloromethane (labeled with CCl$_4$) while the second set includes interactions with chloromethane (labeled as CH$_3$Cl).

In Figure \ref{fig:ana_parameters}, we can see a systematic overestimation of $C_6$ coefficients by the ANA$_d$ model, which could be a result of two factors: First, the value of the damping function in the Becke-Johnson scheme \cite{BJ, BJXDM} is related to the dispersion coefficients through Eqs.~(\ref{eq:rc_equation}) and (\ref{eq:rcrit_scaling}). Since all dispersion coefficients are treated as free parameters, compensation effects may arise. Second, the magnitude for higher-order coefficients might be too weak compared to other components. As a result, the relationship assumed in Eq.~(\ref{eq:rc_equation}) might not hold in the same way as for coefficients obtained from DFT densities.
Similarly, a weak overestimation is found for the molecular polarizabilities. As already noted, this may be caused by a false balance between the short-range induction potential ($V^{ct}$) and the long-range model. Inclusion of induced dipoles in the exchange potential and the parameter used for the Thole damping might also play a role.
In addition, it should be noted that topology-based parameters must compensate for damping effects due to the surrounding environment, which are not accounted for but may play an important role \cite{MBD}.

Overall, the results show that the ANA$_d$ model predicts physically meaningful parameters from scratch given a physically motivated functional form. As such, including constraints through the use of predefined functional forms or known (in-)equalities and parameter-relations might be the method of choice to regularize ML models applied to physical problems.
Further improvements may take into account the relationship between polarizabilites and dispersion coefficients with atomic volume ratios, which has been shown in several studies \cite{TSDispersion, MBD, PolarizabilityScaling, AtomicVolumesBecke, HigherOrderDispersionScaling}. Using an independent model to handle polarizabilities and dispersion coefficients could not only introduce sensitivity to the surrounding environment and conformational changes but also reduce the number of fitted parameters and the resulting interdependencies.

\subsection{General Discussion and Outlook}
The presented results attest to the power and feasibility of the proposed approach to use ML models to predict the parameters of a physically motivated functional form of a FF. 
Most importantly, it indicates that a general FF, which performs well on a wide range of applications, might be attainable. 
To explore whether the ML model can truly learn parameters from the atomic environments alone, no baseline parameters were used in the present work. However, introducing such baseline parameters as well as appropriate scaling of parameters may improve the model in the future.
For the IPA model, this could be accomplished by using an existing FF as baseline and tuning the parameters with the ML model.
For the ANA model, inclusion of additional non-fitted parameters such as the atomic volumes could not only introduce conformational sensitivity but also help to reduce the number of parameters that have to be fitted.
The atomic polarizabilities and dispersion coefficients, in particular, are suitable candidates for such a strategy as they can be obtained from atomic volume ratios given an appropriate model \cite{MBD}.
Including further information such as spectral data from experiments or higher-order derivatives from theoretical calculations could also prove fruitful. This might be particularly interesting for the description of intramolecular interactions.

Model accuracy is mostly limited by two factors: First, an implicit description of electrons is only possible if effects due to changes in the electron distribution can be captured accurately. For this, an accurate description of multipoles is required. Improving multipole prediction can likely improve the model performance further. 
Second, the presented ML-based approach crucially depends on the accuracy and availability of reference data. 
Expected improvements in hardware, software, and methodological advances may provide access to highly accurate calculations of condensed-phase properties in the near future.

\section{Conclusion}
A formalism based on automatic differentiation for the parametrization of classical FF was introduced. The proposed formalism cannot only be viewed as the generalization of commonly used FF definitions and parametrization procedures, it also describes the regularization of ML models through physics.
To showcase its strength, the method was applied to the parametrization of an isotropic-pairwise-additive FF (IPA model) as well as a polarizable FF (ANA model). The performance of the models was first assessed on intermolecular interaction energies of dimers in the gas phase. 
Both approaches perform on par or better relative to comparable methods on several common benchmark data sets, while requiring little computational efforts and human intervention. 
Importantly, the models were exclusively parametrized with respect to the PES of reference QM calculations, without the addition of experimental data.

The IPA$_{d+c}$ model was furthermore applied to the calculation of condensed-phase properties (lattice energies and pure liquid properties), while the ANA$_d$ model was validated by comparing molecular polarizabilities and dispersion coefficients to experimental values.
We found that the IPA$_{d+c}$ model, i.e., a fixed-charge FF parametrized on intermolecular potentials from DFT calculations, can provide consistent results over a wide range of systems, ranging from dimer interaction potentials in vacuum to pure liquid properties. 
While a completely classical description of molecular forces has clear limitations (e.g., no bond formation or breaking), it might still be highly competitive to semi-empirical methods as shown for the ANA model.
In particular, the implicit description of the electron distribution through atomic multipoles offers a very promising alternative to semi-empirical models, providing a comparable level of accuracy but only requiring a fraction of the computational cost.

Automatic differentiation presents a powerful tool for the development of parametrized models, which could also be applied to problems other than molecular interactions.
As shown, this approach accelerates and simplifies the parametrization process of classical FF and can take advantage of large data sets.
In combination with ML based techniques, such as the presented GNN-based atom-typing, the best of both worlds can be harvested. A universal optimisation toolkit combined with robust and physically-constrained models.
In future work, the exploration of additional FF terms and the application of the ANA model to condensed-phase systems will be investigated. In addition, employing the end-to-end differentiable approach to the parametrization of bonded/intramolecular interactions will be explored.

\section{Acknowledgment}
This research was supported by the NCCR MARVEL, a National Centre of Competence in Research, funded by the Swiss National Science Foundation (grant number 182892).

\section{Software and Data Availability}
The intermolecular potentials for $11'489$ molecular crystals which were used during training of the IPA$_{d+c}$ model are available in the ETH research collection (\url{https://doi.org/10.3929/ethz-b-000549359}). 
A GitHub repository including weights and models used to produce the results in this work can be accessed under the following link: \url{https://github.com/rinikerlab/GNNParametrizedFF}.


\renewcommand{\thesection}{A\arabic{section}}
\renewcommand{\thetable}{A\arabic{table}}
\renewcommand{\thefigure}{A\arabic{figure}}
\renewcommand{\theequation}{A\arabic{equation}}
\setcounter{section}{1}
\setcounter{table}{0}
\setcounter{figure}{0}
\setcounter{equation}{0}

\section{Appendix}
\subsection{Potential-Energy Terms: ANA Model}\label{sec:appendix_ana}

\subsubsection{Electrostatics}\label{sec:aniso_es}
The electrostatic interaction is described through the use of multipoles up to quadrupoles. Multipoles were in all cases obtained from our previous equivariant GNN \cite{MultipolesMe} for the prediction of atomic multipoles, which was trained on MBIS reference data \cite{MBIS}. The implementation follows the formalism proposed by Refs.~\cite{MultipoleMethod, MultipoleMethodSmith, MultipoleMethodLin}.
Following the formalism introduced in Ref.~\cite{MultipoleMethod}, the total potential energy due to the interaction of point multipoles at site $i$ and site $j$ is obtained as
\begin{equation}
     V^\text{multi}_{ij}=\sum_{l=0}^4 B_l(r_{ij})G^l(\vec{r}_{ij})
\end{equation}
with $\vec{r}_{ij} = \vec{r}_i - \vec{r}_j$ and $r_{ij} = | \vec{r}_{ij} |$. The radial functions $B_l(r)$ are defined as $B_l(r)=(2l - 1)!!/r^{2l + 1}$  and the coefficients $G^l(\vec{r})$ arise through the interactions between components of two multipole sites. For the considered case of a treatment up to quadrupoles, terms up to $l=4$ are included.
Coefficients $G^l(\vec{r})$ can be thought of as contributions due to the interactions of multipoles of a given order. A list of $G^l(\vec{r})$ is given in Ref.~\cite{MultipoleMethod}.

Deficiencies of the multipole description at short ranges are compensated through the use of the charge penetration model introduced in Ref.~\cite{AMOEBAChargeDamping}.
In this model, effects of interactions between charge distributions are treated through the use of damping functions. Specifically, a damping function for the interaction between a charge distribution and a point charge
\begin{equation}
\label{eq:damp_singlesite}
    f^\text{damp}(r_{ij}) = 1-\exp(-b r_{ij})
\end{equation}
and a damping function for the interaction between two charge distributions
\begin{equation}
\label{eq:damp_overlap}
f^\text{overlap}(r_{ij}) = 1 - \frac{b_j^2}{b_j^2-b_i^2}\exp(-b_i r_{ij}) - \frac{b_i^2}{b_i^2-b_j^2}\exp(-b_j r_{ij})
\end{equation}
are introduced. The damping parameter $b_i$ describes the extent of an exponentially decaying charge distribution centered on atom $i$.
The above damping functions give rise to damping coefficients of a given order $\lambda_l(r)$ which are given in the Supporting Information of Ref.~\cite{AMOEBAChargeDamping}.
Combining the damping coefficients with the radial functions $B_l(r)$ gives rise to the damped radial functions
\begin{equation}
    B_{l}^\text{damp}(r) = \lambda_l(r)B_l(r)
\end{equation}
In the charge penetration model, the standard radial functions are replaced with the damped radial functions. In addition, core-core, core-multipole, and multipole-core interactions are included.
Thus, the complete description of the electrostatic potential is obtained as
\begin{eqnarray} \nonumber
    V^\text{ele}(r_{ij})&=&B_0 z_iz_j + 
    \sum_{l=0}^2 B_{l}^\text{damp}(r_{ij})G_{CM}^l(\vec{r}_{ij})+
    \sum_{l=0}^2 B_{l}^\text{damp}(r_{ij})G_{MC}^l(\vec{r}_{ij})\\
    &+&
    \sum_{l=0}^4 B_{l}^\text{overlap}(r_{ij})G_{MM}^l(\vec{r}_{ij})
\end{eqnarray}
with $z_i$ representing the core charge of atom $i$. 
$B_{l}^\text{damp}(r)$ and $B_{l}^\text{overlap}(r)$ label the aforementioned damping coefficients for a single site and for a pair of charge distributions, respectively.
$G_{CM}^l$, $G_{MC}^l$ and $G_{MM}^l$ describe the core-multipole, multipole-core, and multipole-multipole interactions, respectively.
The core-multipole and multipole-core coefficients are obtained by replacing the monopole with the respective core charge.

\subsubsection{Dispersion}
The dispersion interaction is described based on the formalism used in the XDM model proposed by Becke and Johnson \cite{BJ},
\begin{equation}
    V^\text{disp}(r_{ij}) = -\sum_{n=6, 8, 10}f_n(r_{ij})\frac{C_{n}(i,j)}{(r_{ij})^n}
\end{equation}
with dispersion coefficients $C_{n}$ and a damping function 
\begin{equation}
    f_n(r)=\frac{(r)^n}{(R^\text{vdW})^n + (r)^n}
\end{equation}
depending on a damping parameter $R^\text{vdW}$, which is obtained as follows \cite{BJXDM},
\begin{equation}
    R^\text{vdW} = a_1 R_{c} + a_2 
    \label{eq:rcrit_scaling}
\end{equation}
with two positive parameters $a_1$ and $a_2$ and
\begin{equation}
    R_{c}=\frac{1}{3}\Bigg[\Bigg(\frac{C_{8}}{C_{6}}\Bigg)^\frac{1}{2} +
    \Bigg(\frac{C_{10}}{C_{6}}\Bigg)^\frac{1}{4} +
    \Bigg(\frac{C_{10}}{C_{8}}\Bigg)^\frac{1}{2}\Bigg] .
    \label{eq:rc_equation}
\end{equation}
In the present work, $a_1$ was set to $1$ and $a_2$ to $0$, i.e. $R^\text{vdW} = R_c$.

\subsubsection{Induction}
Polarization is treated based on the Applequist model \cite{Applequist} including the modification proposed by Thole \cite{Thole} and follows the formalism described by Stone \cite{Stone}.
The long-ranged component is treated on the basis of atomic dipoles, which are induced by the external electric field and scaled by the atomic polarizability \cite{Stone}
\begin{equation}\label{eq:polarize}
     M^{(1), \text{ind}} = B^{-1}F^D
\end{equation}
where $F^D$ gathers the electric field components at each atom formed by the static multipoles of the surrounding molecules, i.e., only intermolecular contributions.
The polarizability matrix $B$ is formed as \cite{Stone}
\begin{equation}
    B = \begin{cases}
                   \alpha^{-1}_{ij} & \text{for } i = j\\
                   -T_{ij} & \text{for } i \neq j\\
                    \end{cases}
\end{equation}
with the atomic polarizability $\alpha_i$ and the elements $T_{ij}$ of the dipole-dipole interaction matrix. The $3N\times 3N$ polarizability matrix is inverted to obtain the induced dipoles. Given self-consistently induced dipoles, the potential energy due to induction is given as the inner product with the external field \cite{Stone}
\begin{equation}
    V^{ind}= -\frac{1}{2}M^{(1), \text{ind}} F^D
\end{equation}

To prevent the divergence of induced dipoles ('polarization catastrophe'), elements of the polarizability matrix $B$ are damped based on the modifications proposed by Thole analogously to the damping function used to model charge penetration effects \cite{Thole}.

The exponential damping function as used in AMOEBA was used for this purpose \cite{AMOEBA}
\begin{equation}
    f^\text{Thole}(r)=1-\exp(-au^3(r))
\end{equation}
using a damping factor $a$ and the polarizability-normalized distance
\begin{equation}
    u(r)=\frac{r}{(\alpha_i\alpha_j)^{\frac{1}{6}}}
\end{equation}
As in the original AMOEBA FF, the damping factor $a$ was globally set to $0.39$  \cite{AMOEBA}. 
In addition, a charge transfer potential was added to improve the treatment of polarization at short ranges. This potential is based on work proposed for the AMOEBA+ FF \cite{AMOEBA+} using an exponential form
\begin{equation}\label{eq:charge_transfer}
    V^{ct}(r_{ij})=-A\exp(-Cr_{ij})
\end{equation}
where $A$ describes the strength of the interaction and $C$ is used to approximate the degree of electron density overlap between the respective atom pair.

\subsubsection{Exchange}
The exchange interaction is treated with the anisotropic repulsion model proposed by Rackers \textit{et al.} \cite{AnisotropicExchange}. Using atomic multipoles, their work derives a description for the overlap between two atoms analogously to the electrostatic interaction between atomic multipoles leading to the following expression consistent with the expression obtained by Salem \cite{Salem}.
Specifically, the damping function used to construct the damped radial functions $B_{l}^\text{damp}$ as shown in Section \ref{sec:aniso_es} is replaced with the following damping function 
\begin{equation}\label{eq:damp_exchange}
    f^\text{damp}(r_{ij}) = \frac{\sqrt{r_{ij}}}{b^3}\Bigg(1 + \frac{b r_{ij}}{2} + \frac{1}{3}\bigg(\frac{b r_{ij}}{2}\bigg)^2   \Bigg)\exp{\bigg(\frac{-b r_{ij}}{2}\bigg)}
\end{equation}
for the case $b_i = b_j$ and 
\begin{equation}\label{eq:damp_exchange_pair}
    f^\text{damp}(r_{ij}) = \frac{1}{2X^3\sqrt{r_{ij}}}
    \Bigg(b_i(rX-2b_j)\exp{\bigg(\frac{-b_j r_{ij}}{2}\bigg)}
    +b_j(rX+2b_i)\exp{\bigg(\frac{-b_i r_{ij}}{2}\bigg)}\Bigg)
\end{equation}
for the case $b_i \neq b_j$. With $X=\big(\frac{b_i}{2}\big)^2 - \big(\frac{b_j}{2}\big)^2$
leading to the radial function for the exchange potential
\begin{equation}\label{eq:damp_B0}
    B_{0}^\text{damp}(r_{ij})=\frac{b_i^3b_j^3}{r_{ij}}f^\text{damp}(r_{ij})^2
\end{equation}
with higher order radial functions $B_{0}^\text{damp}(r)$ following analogously to the damped radial functions presented in the description of the electrostatic potential. 

The overlap defined as
\begin{equation}
    S_\text{total}^2 = \sum_{l=0} B_{l}^\text{damp}(r)G^l(\vec{r})
\end{equation}
is then used to obtain the exchange potential energy contribution,
\begin{equation}\label{eq:exchange}
    V^{ex}(r_{ij}) =\frac{k_ik_j}{r_{ij}}S^2
\end{equation}
with $k_i$ being the relative size of atom $i$, and $S^2$ the multipole derived orbital overlap. 

We note that $b$ used to damp the exchange interaction (Eqs.~(\ref{eq:damp_exchange}) and (\ref{eq:damp_exchange_pair})) and the electrostatic interaction (Eqs.~(\ref{eq:damp_singlesite}) and (\ref{eq:damp_overlap})) are treated as independent parameters despite relating to the same underlying feature, i.e., an exponentially decaying charge distribution. Induced dipoles are added to the static dipoles.
As in the original work \cite{AnisotropicExchange}, the monopole is replaced with an additional atomic parameter $q^\text{val}$, which weights the influence of the multipole interaction coefficients present in $G^l(\vec{r})$. Following Ref.~\cite{AnisotropicExchange}, $q^\text{val}$ is set to $1$ for all hydrogens and $>2$ for all other elements. This parameter is added to the negative monopole, yielding the final $q^{ex}$ parameter, which is used in place of the monopole used to compute the multipole interaction coefficients in $G^l(\vec{r})$.

\subsection{Graph Construction}\label{sec:appendix_graph_construction}
Since topological information is not available for all data sets, graphs were constructed from monomer coordinates. Each graph was built by adding a node for each atom and an edge between bonded nodes. Bonds were added by first assigning hydrogen and halogen atoms to its nearest neighbours. For all other elements, all nearest neighbours within a given cutoff were assigned as bonded neighbours to the respective central atom. For C, N, O and S, a cutoff of $2.0$\,\AA, $1.8$\,\AA, $1.8$\,\AA\, and $2.25$\,\AA\,, respectively, was used. 
Element types were encoded as one-hot vectors serving as node features. Edge features were built by concatenating the node features of the binding partners. 
We note that no distance information or chemical concepts, such as bond types, were included in the graph except for the aforementioned assumptions regarding the extraction of bonded neighbours. 

\subsection{GNN}\label{sec:appendix_gnn_model}
Following the previously introduced notion, the parametrization model consists of a GNN and a readout layer or combination-rule layer.
Node and edge features were initially embedded as $64$-dimensional vectors. The GNN consisted of independent graph-updating layers, which were composed of two fully connected feed-forward layer with $64$ units, each combined with the Mila non-linearity using $\beta=-1$ \cite{Mila}.

Each edge- and node update layer consisted of the following module ($[64, \text{Mila}, 64, \text{Mila}]$) for the IPA model and ($[64, \text{Mila}]$) for the ANA model.
For both models, three graph-updating layers were used.
The GNN module was followed by a readout module/combination rule parametrized by two fully connected layers and an output layer with $n$ output neurons equivalent to the number of predicted parameters $[64, \text{Mila}, 64, \text{Mila}, \text{n}, \text{Softplus}+\epsilon]$.
For the ANA model, a small term ($\epsilon=10^{-3}$) was added to the output of the Softplus activation to avoid numerical instabilities, and two independent readout modules were used.
Layer weights were initialized with the method introduced by He \cite{HE}. 

GNN models and intermolecular potentials were implemented with TensorFlow (2.6.2) \cite{TensorflowPaper, TensorflowSoftware} and the GraphNets library (1.1.0) \cite{GNN} using the InteractionNetwork model \cite{InteractionNetwork}. 
Pipelines were written with Python (3.9.5) \cite{Python} and Numpy (1.19.5) \cite{Numpy}.
Plots and visualizations were created with Matplotlib (3.5.1) \cite{PLT} and Seaborn (0.11.2) \cite{Seaborn}.
Trajectories were processed and analyzed with MDTraj (1.9.7) \cite{MDTraj}.
RDKit (2021.09.2) was used to manipulate molecules and generate conformations \cite{RDKIT, ETKDG}.

\subsection{Model Optimization}
Model parameters were optimized using ADAM and the same exponential decay schedule used for annealing with learning rates ($4\cdot 10^{-4},\, 4\cdot 10^{-6}$) \cite{ADAM}.
Gradients were clipped by their global norm with a clip norm of $1$ \cite{GradientClipping}.

\subsection{Condensed-Phase Simulations}\label{sec:appendix_simulation_protocol}
Condensed-phase simulations as well as the evaluation of the electrostatic potential for intermolecular potentials of crystals were performed under periodic boundary conditions using OpenMM (7.7) \cite{OpenMM7}. 
Bonded terms were parametrized with OpenFF 2.0 \cite{SageFF} as our model does currently not provide bonded terms. The C6-C9 potential was implemented using the CustomNonbondedForce class in OpenMM. 

For the IPA$_{d+c}$ model, monopoles were predicted for each conformation of an ensemble generated with the ETKDG conformation generator \cite{ETKDG} implemented in the RDKit \cite{RDKIT}.
A RMS pruning threshold of $0.1\,$\AA\, was used, and up to $32$ conformations were generated. Monopoles were predicted for each conformation
using our previously introduced equivariant GNN \cite{MultipolesMe}. Fixed partial charges were then obtained by averaging over all monopoles obtained for the conformational ensemble.
Charges remained fixed during the simulation, and the same charges were used for the condensed-phase as well as the vacuum simulations.
The 1,4-electrostatic interactions were scaled with the same factor as in OpenFF.
The 1,4-Lennard-Jones interactions were described with the C$_6$--C$_12$ parameters from OpenFF 2.0, and a scaling factor of $0.5$ was used.
Long-range electrostatics beyond the cutoff were treated with the smooth particle-mesh-Ewald (PME) method \cite{PME2}. C$_6$--C$_9$ terms were included up to a distance of $10\,$\AA. No long-range correction or shifting function was used for the van der Waals interactions.

Initial configurations were generated using packmol \cite{Packmol} and conformations generated with ETKDG \cite{ETKDG, RDKIT}. 
The number of molecules was chosen such that a cubic box with side lengths $50\,$\AA~at the experimental density would be filled. 
Configurations were sampled from an NPT ensemble at $298.15\,$K using a Langevin integrator \cite{LangevinIntegrator} with a time step of $2\,$fs and a collision frequency of $1\,$ps$^{-1}$.
To maintain constant pressure, a Monte Carlo barostat \cite{MCBarostat} with a target pressure of 1~bar and a trial move every $25$th step was applied to the system.. 
Bonds with hydrogen atoms were fixed at the equilibrium distance with the LINCS algorithm \cite{LINCS}.

Each box was equilibrated for $2\,$ns followed by a $20\,$ns production run, with system data being saved to disk every $4\,$ps. System properties were averaged over the whole production run. 
To obtain the potential energy in the gas phase, a single molecule was simulated in vacuum using the same settings and simulation times as for the condensed-phase simulation.

\subsection{Training Data}\label{sec:appendix_trainingsets}
In the following sections we provide an overview of the data sets used.
Generally, only neutral molecules with elements included in $\{H, C, N, O, F, S, Cl\}$ were used. 

\subsubsection{Data Set I: Dimers in Vacuum}\label{sec:appendix_dimerset}
All models were initially fitted to a recently published DES5M data set of small-molecule dimer dissociation curves published by Donchev \textit{et al.} \cite{DEShawDimers}. 
Both, the SAPT0 components \cite{SAPT0, SAPT1} and the spin-network-scaled-MP2 \cite{SNSMP2, SNS, MP2} total intermolecular potential were used during training.

\subsubsection{Data Set II: Crystal Intermolecular Potentials}\label{sec:appendix_crystal_data}
To fit intermolecular potentials in the crystalline phase, a new data set with DFT energies of molecular crystals was built.
Calculations for this data set were performed on the Euler cluster of ETH Zürich.
Experimental crystal structures from the CSD, which satisfy the following requirements were selected:
\begin{enumerate}
    \item A single molecule in the asymmetric unit
    \item Up to $100$ atoms in the unit cell 
    \item No disorder or missing coordinates
    \item Unit-cell volume up to $1600$\,A$^3$
    \item Elements in $\{H, C, N, O, F, Cl, S\}$
\end{enumerate}
A total of $35'577$ structures were found to satisfy these requirements. Of which $32'811$ were successfully relaxed in less than $192$\,CPU hours under the following settings. 
Structures were relaxed with the L-BFGS optimizer using the plane wave code QuantumEspresso (QE, 6.8) under the PBE functional and XDM dispersion correction \cite{QE1, QE2, QE3, PBE, PBE1, BJXDM, XDM}. PBE in combination with the XDM dispersion correction was shown to perform well for molecular crystals \cite{XDMB86bPBE}. 
Coordinates were relaxed under fixed lattice parameters with default QE settings, i.e. an energy convergence threshold of $10^{-4}$\,Ry and a force convergence threshold of $10^{-3}$\,a.u. The plane wave cutoff was set to $70$\,Ry and the charge density cutoff to $560$ Ry. A uniformly spaced k-point grid was used with the number of k-points for each dimension chosen such that $n_{k_i} = \ceil*{\frac{25}{|x_i|}}$.
Cutoffs were chosen such that the lattice energy of the optimized structures of a balanced subset of the X23 database (ACETAC07, ANTCEN13, CYTSIN01, ECARBM01, HXMTAM10, TRIZIN, TROXAN11, URACIL) was converged to less than $10^{-2}$\,kJ/mol per atom \cite{X23Johnson, X23Johnson}. Further, the publicly available projector augmented-wave (PAW) pseudopotentials (PP) \cite{PSLibrary} were used.
In comparison with other publicly available PP, they provided the most accurate results during our convergence studies. The accuracy was determined as the MAE of the lattice energy with respect to experimental values.
For a subset of $11'666$ minimized structures, five additional geometries were generated by scaling the unit cell by factors of (0.95, 0.975, 1.0, 1.05, 1.1) without modifying the asymmetric unit. For each structure, a single-point calculation was performed using the XDM dispersion corrected B86BPBE functional \cite{PBE, B86}. Previous work has shown that B86BPBE-XDM reproduces energies with very high accuracy \cite{XDMB86bPBE}. PAW PP for the B86BPBE functional were generated with the pslibrary (1.0) \cite{PSLibrary}. The same k-points scheme was used, while the plane wave cutoff and the charge density cutoff were set to $80$\,Ry and $800$\,Ry, respectively. Monomers were calculated with a single k-point sampled at $\Gamma$ in a cubic box with lengths chosen such that the minimal distance between atoms of the central cell and its periodic images was larger then $12$\,\AA.

The intermolecular potential energies used to train the IPA$_{d+c}$ model were then obtained as
\begin{equation}
    \Delta V_\text{inter} = \frac{V_{uc}}{Z} - V_{g}
\end{equation}

The crystal dataset is available in the ETH research collection (\url{https://doi.org/10.3929/ethz-b-000549359}).

\printbibliography

\end{document}